\documentstyle[aps,prd,eqsecnum,graphicx]{revtex}

\begin{document}

\draft

\title{Quantum analysis of Rydberg atom cavity detector
\protect \\
for dark matter axion search}

\author{Akira Kitagawa,%
\footnote{E-mail address: kitagawa@nucleng.kyoto-u.ac.jp}
Katsuji Yamamoto%
\footnote{E-mail address: yamamoto@nucleng.kyoto-u.ac.jp}}
\address{Department of Nuclear Engineering,
Kyoto University, Kyoto 606-8501, Japan}

\author{Seishi Matsuki%
\footnote{E-mail address: matsuki@carrack.kuicr.kyoto-u.ac.jp}}
\address{Nuclear Science Division, Institute for Chemical Research,
Kyoto University, Uji, Kyoto 611-0011, Japan}


\maketitle

\begin{abstract}
Quantum calculations are developed on the dynamical system consisting
of the cosmic axions, photons and Rydberg atoms which are interacting
in the resonant microwave cavity.
The time evolution is determined for the number of Rydberg atoms
in the upper state which are excited
by absorbing the axion-converted and thermal background photons.
The calculations are made, in particular, by taking into account
the actual experimental situation such as the motion and uniform
distribution of the Rydberg atoms in the incident beam
and also the spatial variation of the electric field in the cavity.
Some essential aspects on the axion-photon-atom interaction
in the resonant cavity are clarified by these detailed calculations.
Then, by using these results the detection sensitivity
of the Rydberg atom cavity detector is estimated properly.
This systematic quantum analysis enables us to provide
the optimum experimental setup for the dark matter axion search
with Rydberg atom cavity detector.
\end{abstract}


\section{Introduction}
\label{sec:intro}

Search for the so-called ``invisible" axions
\cite{Peccei77,WWBT78,KSVZ,DFSZ,KS85}
as non-baryonic dark matter is one of the most challenging issues
in particle physics and cosmology
\cite{Kolb90,Smith90}.
The mass range $ m_a \sim 10^{-6}{\rm eV} - 10^{-3}{\rm eV} $
is still open for the cosmic axions
\cite{Kolb90,Raffelt90}.
(The light velocity is set to be $ c = 1 $ throughout the present article.)
As originally proposed by Sikivie
\cite{S83K85},
the basic idea for the dark matter axion search is to convert axions
into microwave photons in a resonant cavity under a strong magnetic field
via the Primakoff process.
It is, however, extremely difficult to detect the cosmic axions
due to their unusually weak interactions with ordinary matters.
Pioneering tries with amplification-heterodyne-method
were published already
\cite{dePWH}.
An advanced experiment by the US group is currently continued,
and some results have been reported,
where the KSVZ axion of mass $ 2.9 \times 10^{-6} {\rm eV} $
to $ 3.3 \times 10^{-6} {\rm eV} $ is excluded at the 90\% confidence level
as the dark matter in the halo of our galaxy
\cite{Hagmann98}.

In this paper, we describe a quite efficient scheme
for the dark matter axion search,
where Rydberg atoms are utilized to detect the axion-converted
microwave photons
\cite{MY91,MOY94,OMY96,YM99,TKKSFYMM99}.
An experimental apparatus called CARRACK I
({\underline C}osmic {\underline A}xion {\underline R}esearch
with {\underline R}ydberg {\underline A}toms
in resonant {\underline C}avities in {\underline K}yoto)
is now running to search for the dark matter axions
over a 10 \% mass range around $ 10^{-5} {\rm eV} $.
Based on the performance of CARRACK I, a new large-scale apparatus
CARRACK II has been constructed recently
to search for the axions over a wide range of mass
\cite{TKKSFYMM99}.
Clearly, in order to derive quantitative and rigorous results
from the axion search with this type of Rydberg atom cavity detectors,
it is essential to develop the quantum theoretical formulations
and calculations on the interaction of cosmic axions with Rydberg atoms
via the Primakoff process in the resonant cavity
\cite{OMY96,YM99}.
Here, we present the details of these quantum calculations.
This theoretical analysis actually provides important guides
for the detailed design of Rydberg atom cavity detector,
which will be examined separately in forthcoming papers.
It is also fascinating that these quantum theoretical investigations
on the axion-photon-atom interaction will be useful
in viewpoint of the applications of the cavity quantum electrodynamics
to fundamental researches.

This paper is organized as follows.
In Sec. \ref{sec:detector}, we describe the Rydberg atom cavity detector
for the dark matter axion search.
In the quantum theoretical point of view,
this Rydberg atom cavity detector can be treated as the dynamical system
of interacting oscillators with dissipation.
In Sec. \ref{sec:system}, these quantum oscillators
describing the photons, axions and atoms are introduced appropriately.
Then, in Sec. \ref{sec:interactions},
the interaction Hamiltonians are provided
in terms of these quantum oscillators.
In Sec. \ref{sec:evolution},
the quantum dynamics of interacting oscillators
with dissipation is generally formulated in the Liouville picture.
A series of master equations is derived
to determine the time evolution of the quatum averages
of the occupation numbers and higher-order correlations
of these oscillators.  By applying this formulation
to the axion-photon-atom system in the resonant cavity,
we can, in particular, calculate the number of the atoms in the upper state
which are excited by absorbing the axion-converted
and thermal background photons.
In Sec. \ref{sec:aspects}, we examine some characteristic properties
of the axion-photon-atom interaction by solving analytically
the master equation for the simple case with time-independent
atom-photon coupling.
Then, in Sec. \ref{sec:sensitivity},
in order to make precise estimates on the sensitivity
of the Rydberg atom cavity detector, we elaborate
the calculations by taking into account the motion
and uniform distribution of the Rydberg atoms in the incident beam
and also the spatial variation of the electric field in the cavity.
In Sec. \ref{sec:dependence}, the dependence of the detection sensitivity
on the relevant experimental parameters is discussed.
Detailed numerical calculations are performed in Sec. \ref{sec:numerical},
and the quantum evolution of the axion-photon-atom system in the resonant
cavity is determined precisely.
The counting rates of signal and noise are calculated
with the steady solutions of the master equation.
The sensitivity of the Rydberg atom cavity detector
is then estimated from these calculations.
Finally, we summarize the present quantum analysis
in Sec. \ref{sec:sammary}.
Appendices are devoted to some supplementary issues.

\section{Rydberg atom cavity detector for dark matter axion search}
\label{sec:detector}

\begin{figure}
\begin{center}
\scalebox{1.2}[1.2]{\includegraphics*[6cm,14cm][16cm,27cm]{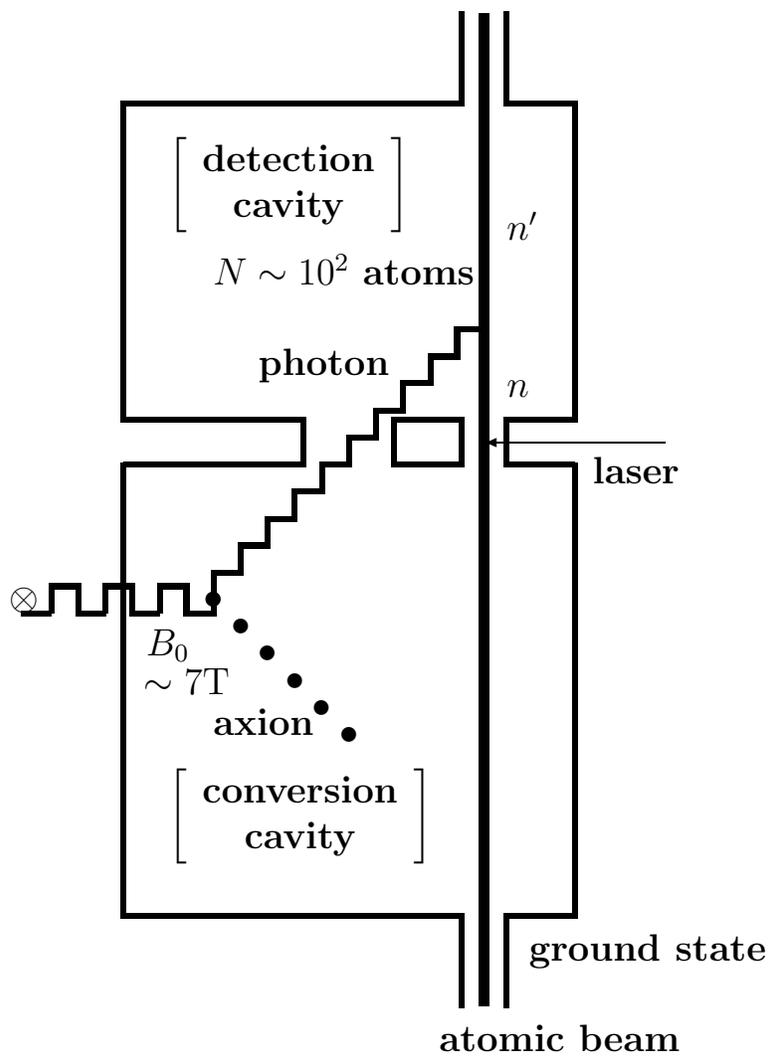}}
\caption{A schematic diagram of the Rydberg atom cavity detector.
\label{fig:scheme}}
\end{center}
\end{figure}

A schematic diagram of the Rydberg atom cavity detector
(CARRACK I) is shown in Fig. \ref{fig:scheme}.
The axions are first converted into photons under a strong magnetic field
in the conversion cavity.  Then, the photons are transferred
to the detection cavity via a coupling hole, and they are absorbed there
by Rydberg atoms. The detection cavity is set to be free from  magnetic
field to avoid the complexity of the atomic energy levels
due to the Zeeman splitting.  The Rydberg atoms,
the transition frequency of which is tuned approximately
to the cavity resonant frequency $ \sim 1 {\rm GHz} $,
are prepared initially in a lower state
with principal quantum number $ n \sim 100 $
by exciting alkaline atoms in the ground state with laser excitation
just in front of the detection cavity.  It should, however, be noted
that atoms in the upper state with $ n^\prime \ ( > n ) $ are not generated
at this stage.
The atoms prepared in this way are excited to the upper state
by absorbing the microwave photons in the detection cavity,
and they are detected quite efficiently
with the selective field ionization method
\cite{Gallagher94}
just out of the cavity.
By cooling the resonant cavity system down to about 10 mK
with a dilution refrigerator in high vacuum,
the thermal background photons can be reduced
sufficiently to obtain a significant signal-to-noise ratio.
Hence the Rydberg atom cavity detector,
which is free from the amplifier noise by itself,
is expected to be quite efficient for the dark matter axion search.

In the quantum theoretical point of view,
the Rydberg atom cavity detector can be treated
as the dynamical system of the interacting oscillators with dissipation
which appropriately describe the axions, photons and Rydberg atoms.
In the following sections we develop detailed quantum theoretical
formulations and calculations for this dynamical system.
Specifically, the analysis is made by taking into account
the actual experimental situation such as the motion and uniform distribution
of the Rydberg atoms in the incident beam
and also the spatial variation of the electric field in the cavity.
This quantum treatment provides proper estimates
on the sensitivity of the Rydberg atom cavity detector
for dark matter axion search.

\section{Axion-photon-atom system in the cavity}
\label{sec:system}

We first identify the quantum oscillators which appropriately
describe the photons, axions and Rydberg atoms in the resonant cavity.
This provides the basis for investigating
various properties of the axion-photon-atom system.

\subsection{Resonant mode of photons}

The electric field operator in the cavity $ {\cal V} $ is given by
\begin{equation}
{\mbox{\boldmath $ E $}} ({\bf x},t)
= ( \hbar \omega_c / 2 \epsilon_0 )^{1/2}
[ {\mbox{\boldmath $ \alpha $}} ({\bf x}) c(t)
+ {\mbox{\boldmath $ \alpha $}}^* ({\bf x}) c^{\dagger}(t) ]
\label{eqn:E}
\end{equation}
for the radiation mode with a resonant frequency $ \omega_c $,
where $ \epsilon_0 $ is the dielectric constant.
The creation and annihilation operators satisfy
the usual commutation relation
\begin{equation}
[ c , c^\dagger ] = 1 .
\label{eqn:ccdgg}
\end{equation}
The mode vector field $ {\mbox{\boldmath $ \alpha $}}({\bf x}) $
is normalized by the condition
\begin{equation}
\int_{\cal V} | {\mbox{\boldmath $ \alpha $}} ({\bf x}) |^2 d^3 x = 1 .
\label{eqn:a-nrm}
\end{equation}
The whole cavity $ {\cal V} $
may be viewed as a combination of two subcavities,
the conversion cavity $ {\cal V}_1 $ with volume $ V_1 $
and the detection cavity $ {\cal V}_2 $ with volume $ V_2 $,
which are coupled together:
\begin{equation}
{\cal V} = {\cal V}_1 \oplus {\cal V}_2 .
\end{equation}
The axion-photon conversion takes place in $ {\cal V}_1 $
under the strong mangetic field,
while the Rydberg atoms are excited by absorbing the photons
in $ {\cal V}_2 $.  It is then suitable to divide the mode vector as
\begin{equation}
{\mbox{\boldmath $ \alpha $}} ({\bf x})
= {\mbox{\boldmath $ \alpha $}}_1 ({\bf x})
+ {\mbox{\boldmath $ \alpha $}}_2 ({\bf x}) ,
\label{eqn:modev}
\end{equation}
where $ {\mbox{\boldmath $ \alpha $}}_1 ({\bf x}) = {\bf 0} $
for $ {\bf x} \in {\cal V}_2 $
and $ {\mbox{\boldmath $ \alpha $}}_2 ({\bf x}) = {\bf 0} $
for $ {\bf x} \in {\cal V}_1 $, respectively.
The normalization condition of $ {\mbox{\boldmath $ \alpha $}} ({\bf x}) $
is rewritten as
\begin{equation}
\int_{{\cal V}_1} | {\mbox{\boldmath $ \alpha $}}_1 ({\bf x}) |^2 d^3 x
+ \int_{{\cal V}_2} | {\mbox{\boldmath $ \alpha $}}_2 ({\bf x}) |^2 d^3 x
= 1 .
\label{eqn:moden}
\end{equation}

The actual cavity is designed so that neglecting the small joint region
the subcavities $ {\cal V}_1 $ and $ {\cal V}_2 $ admit the mode vectors
$ {\mbox{\boldmath $ \alpha $}}_1^0 ({\bf x}) $
and $ {\mbox{\boldmath $ \alpha $}}_2^0 ({\bf x}) $
(up to the normalization and complex phase), respectively,
whose frequencies are tuned to be almost equal to some common value
$ \omega_c^0 $.
In this situation, as confirmed by numerical calculations
and experimental observations,
two nearby eigenmodes with the frequencies
$ \omega_c ,\ \omega_c^\prime \simeq \omega_c^0 $ are obtained
for the whole cavity $ {\cal V} $.  Then, the mode vector
$ {\mbox{\boldmath $ \alpha $}} ({\bf x}) $ is constructed
approximately of
$ {\mbox{\boldmath $ \alpha $}}_1 ({\bf x})
\simeq {\mbox{\boldmath $ \alpha $}}_1^0 ({\bf x}) $
and $ {\mbox{\boldmath $ \alpha $}}_2 ({\bf x})
\simeq {\mbox{\boldmath $ \alpha $}}_2^0 ({\bf x}) $
with significant magnitudes in both $ {\cal V}_1 $ and $ {\cal V}_2 $.
The conversion of the cosmic axions takes place predominantly
to the radiation mode which is resonant with the axions
satisfying the condition
$ | \omega_c - m_a / \hbar | \lesssim \gamma_a $
(axion width) $ \sim $ small fraction of $ \gamma $ (cavity damping rate),
as will be discussed later.
The cavity can be designed so as to give a sufficient separation
of $ | \omega_c - \omega_c^\prime | > {\rm several} \ \gamma $
for the nearby modes with strong coupling
between $ {\cal V}_1 $ and $ {\cal V}_2 $.
Therefore, in the search for the signal from the cosmic axions,
the one resonant mode can be extracted solely
for the electric field in a good approximation, as given
in Eq. (\ref{eqn:E}), whose frequency $ \omega_c $ is supposed to be
close enough to the axion frequency $ \omega_a = m_a / \hbar $.

\subsection{Coherent mode of cosmic axions}

The axion field operator is expanded as usual
in terms of the continuous modes:
\begin{eqnarray}
\phi ({\bf x},t) & = & \hbar^{1/2} \int \frac{d^3 k}{( 2 \pi )^3 2 \omega_k}
\left[ a_{\bf k} (t) {\rm e}^{i{\bf k}\cdot{\bf x}}
+ a_{\bf k}^{\dagger} (t) {\rm e}^{-i{\bf k}\cdot{\bf x}} \right]
\nonumber \\
& & \nonumber \\
& \equiv &  \phi^+ ({\bf x},t) + \phi^- ({\bf x},t) ,
\label{eqn:phi}
\end{eqnarray}
where $ \phi^+ $ and $ \phi^- $ represent the positive and negative
frequency parts, respectively.
The cosmic axions form a coherent state with momentum distribution
$ \eta_a ({\bf k}) $,
\begin{equation}
| \eta_a \rangle = N_\eta^{-1/2}
\exp \left( \int \frac{d^3 k}{( 2 \pi )^3 2 \omega_k}
\eta_a ({\bf k}) a_{\bf k}^{\dagger} \right) | 0 \rangle ,
\label{eqn:eta-a}
\end{equation}
where $ N_\eta $ is the normalization constant to ensure
the condition $ \langle \eta_a | \eta_a \rangle = 1 $.
The velocity dispersion of the galactic axions is expected to be very small
as $ \beta_a \sim 10^{-3} $ so that $ \eta_a ({\bf k}) $
takes significant values only in a small region $ {\cal R}_a $
where $ | {\bf k} | \lesssim \beta_a m_a / \hbar $
\cite{Kolb90,Smith90}.

The axion field operator (\ref{eqn:phi}) may be divided into
the low-momentum part $ \phi_{{\cal R}_a} $
with $ {\bf k} \in {\cal R}_a $ and the residual part $ \phi_{\rm res} $:
\begin{equation}
\phi ({\bf x},t) = \phi_{{\cal R}_a} ({\bf x},t)
+ \phi_{\rm res} ({\bf x},t) .
\label{eqn:phicomp}
\end{equation}
Since the coherent region $ {\cal R}_a $ is chosen so as to give
$ \langle \eta_a | \phi | \eta_a \rangle \approx
\langle \eta_a | \phi_{{\cal R}_a} | \eta_a \rangle $,
the residual part $ \phi_{\rm res} $ does not provide
significant contributions in the following calculations, i.e.,
$ \langle \eta_a | \phi_{\rm res} | \eta_a \rangle \approx 0 $.
It is further noticed that the spatial variation of
$ \phi_{{\cal R}_a}({\bf x},t) $ is negligible in the cavity region,
i.e., $ {\rm e}^{i{\bf k}\cdot{\bf x}} \simeq 1 $
for $ | {\bf k} | \lesssim \beta_a m_a / \hbar $
and $ {\bf x} \in {\cal V}_1 $.
This is because the de Broglie wavelength of the axions,
$ \lambda_a \simeq h /( \beta_a m_a ) \sim 100{\rm m} $
typically for $ m_a \sim 10^{-5} {\rm eV} $ and $ \beta_a \sim 10^{-3} $,
is much longer than the microwave cavity scale
$ \sim 0.1 {\rm m} $.
In this situation to ensure approximately
$ \phi_{{\cal R}_a}({\bf x},t) \approx \phi_{{\cal R}_a}({\bf 0},t) $
in the cavity, the coherent axion mode can be identified as
\begin{eqnarray}
a(t) & = &
( \hbar \Sigma_a )^{-1/2} \phi_{{\cal R}_a}^+ ({\bf 0},t)
\nonumber \\
&=& \Sigma_a^{-1/2}
\int_{{\cal R}_a} \frac{d^3 k}{(2 \pi )^3 2 \omega_k} \ a_{\bf k} (t) .
\label{eqn:a(t)}
\end{eqnarray}
Here, the normalization factor is given by
\begin{equation}
\Sigma_a = \int_{{\cal R}_a} \frac{d^3 k}{(2 \pi )^3 2 \omega_k}
\equiv \frac{1}{2m_a} \left( \frac{\beta_a m_a}{2 \pi \hbar} \right)^3 ,
\label{eqn:sigma-a}
\end{equation}
so that the coherent mode operator satisfies
the canonical commutaion relation,
\begin{equation}
[a, a^{\dagger}] = 1 .
\label{eqn:a-comm}
\end{equation}
The normalization factor $ \Sigma_a $ is expressed in terms of
the mean velocity $ \beta_a $ of galactic axions
with $ \omega_k \simeq m_a / \hbar $ and $ \beta_a \ll 1 $.
It should, however, be realized that this parameter $ \beta_a $
is introduced just for the normalization of the coherent axion mode.
The actual mean velocity of axions is rather determined
by the axion spectrum, which may slightly be different
from the parameter $ \beta_a $ given in Eq. (\ref{eqn:sigma-a}).
The normalization factor $ \Sigma_a $ is canceled out anyway
in calculating the axion signal rate, as explicitly seen later.

The coherent cosmic axions can be described effectively
in terms of a single mode oscillator, as shown in the above.
Then, the energy spread of the galactic axions can be taken into account
as the damping rate of the coherent axion mode,
\begin{equation}
\gamma_a \sim \beta_a^2 m_a / \hbar .
\label{eqn:gam_a}
\end{equation}
The expectation value of the number operator $ a^\dagger a $
of the coherent axion mode is determined by the energy density
$ \rho_a $ of the cosmic axions as follows.
The number density operator for the axion field is given by
$ {\hat n}_a ({\bf x},t) = ( 2 \hbar )^{-1}
(i \phi^- \partial_0 \phi^+ -i \partial_0 \phi^- \phi^+ ) $,
and its expectation value is calculated for $ {\bf x} \in {\cal V}_1 $
and $ \beta_a \ll 1 $ as
\begin{eqnarray}
\langle \eta_a | {\hat n}_a ({\bf x},0) | \eta_a \rangle
& \simeq &
\hbar^{-2} m_a \langle \eta_a | \phi_{{\cal R}_a}^- ({\bf 0},0)
\phi_{{\cal R}_a}^+ ({\bf 0},0) | \eta_a \rangle
\nonumber \\
& \simeq &\rho_a / m_a .
\label{eqn:rhoa-na}
\end{eqnarray}
Then, by considering Eq.(\ref{eqn:a(t)})
to relate the relevant part of axion field operator $ \phi_{{\cal R}_a} $
to the coherent mode operators $ a $ and $ a^\dagger $, we can calculate
the occupation number of the coherent axion mode as
\begin{equation}
{\bar n}_a = \langle \eta_a | a^{\dagger}(0) a(0) | \eta_a \rangle
\simeq \left( \frac{2 \pi \hbar}{\beta_a m_a} \right)^3
\left( \frac{\rho_a}{m_a} \right) .
\label{eqn:na}
\end{equation}
This expression implies that the coherent axion mode
is normalized suitably in a box with a volume
of $ ({\mbox{de Broglie wavelength}})^3 $.

\subsection{Oscillator for Rydberg atoms}

The Rydberg atom is treated well as a two-level system
in the resonant cavity.
The relevant lower and upper atomic states are represented
by $ | n \rangle $ and $ | n^\prime \rangle $, respectively.
They are connected by the electric dipole transition
with frequency $ \omega_b = ( E_{n^\prime} - E_n ) / \hbar $,
which is actually fine-tuned to be almost equal
to the cavity frequency $ \omega_c $ by utilizing the small Stark shift.
This two-level atomic system is described
in terms of spin 1/2 like operators
\cite{Haroche85}
\begin{eqnarray}
D^+ &=& | n^\prime \rangle \langle n | , \
D^- = | n \rangle \langle n^\prime | , \nonumber \\
D^3 &=& \frac{1}{2} \left( \ | n^\prime \rangle \langle n^\prime |
- | n \rangle \langle n | \ \right) .
\label{eqn:D-op}
\end{eqnarray}
These operators satisfy the usual SU(2) Lie algebra,
\begin{equation}
[ D^+ , D^- ] = 2 D^3 , \ [ D^3 , D^{\pm} ] = \pm D^{\pm} .
\label{eqn:D-comm}
\end{equation}
The free Hamiltonian for the two-level system is then given by
\begin{equation}
H_{\rm atom} = \hbar \omega_b D^3 .
\label{eqn:HatomD}
\end{equation}

The Rydberg atoms are initially prepared in the lower state
in the present detectoin scheme for cosmic axions.
It is also expected that the probability of the excitation
of an atom to the upper state by absorbing a photon is small enough.
This is because the average number of photons in the cavity
is much smaller than one at sufficiently low temperatures
($ \sim 10 {\rm mK} $).
Therefore, the atoms almost remain in the lower state,
providing a good approximation
\begin{equation}
D^3 \approx \langle n | D^3 | n \rangle = - \frac{1}{2}
\label{eqn:D3}
\end{equation}
in the first commutation relation of Eq.(\ref{eqn:D-comm}).
Then, the two-level atomic system operators may be substituted
by those of an oscillator as
\cite{Haroche85}
\begin{equation}
b \approx D^- , \ b^\dagger \approx D^+ ,
\label{eqn:b-op}
\end{equation}
which satisfy the commutation relation
\begin{equation}
[ b , b^\dagger ] = 1 .
\label{eqn:b-comm}
\end{equation}
The second commutation relation of Eq.(\ref{eqn:D-comm})
is also reproduced with $ D^3 = - \frac{1}{2} + b b^\dagger $
up to the higher order terms,
and accordingly the free atomic Hamiltonian is expressed effectively as
\begin{equation}
H_{\rm atom} = \hbar \omega_b b^\dagger b ,
\label{eqn:Hatom}
\end{equation}
where the irrelevant constant $ \hbar \omega_b / 2 $ is discarded.
This treatment of the Rydberg atom in terms of the quantum oscillator
is valid if the probability in the upper state (or the lower state)
is very small, i.e., $ \langle b^\dagger b \rangle \ll 1 $,
as is the case in the present scheme.

\subsection{Characteristics of the photons, axions and atoms}

We have seen so far that the resonant photons, coherent cosmic axions
and Rydberg atoms in the resonant cavity are described
in terms of the appropriate quantum oscillators
with dissipation due to the couplings to the relevant reservoirs.
The characteristic properties of these quantum oscillators
are summarized as follows.

The thermal photon number $ {\bar n}_c $ of the resonant mode
is determined by the cavity temperature $ T_c $ as
\begin{equation}
{\bar n}_c
= \left( {\rm e}^{\hbar \omega_c / k_{\rm B} T_c} - 1 \right)^{-1} .
\label{eqn:nbarc}
\end{equation}
The damping rate $ \gamma_c $ of photons is described
in terms of the quality factor $ Q $ of the cavity as
\begin{equation}
\gamma_c \equiv \gamma =  5 \times 10^{-10} {\rm eV} \hbar^{-1}
\left( \frac{\hbar \omega_c}{10^{-5}{\rm eV}} \right)
\left( \frac{2 \times 10^4}{Q} \right) .
\label{eqn:gamma}
\end{equation}
The coherent axion mode is normalized in a box of the de Broglie wavelength,
as seen in Eq. (\ref{eqn:na}).
The axion number is then estimated as
\begin{equation}
{\bar n}_a = 5.7 \times 10^{25}
\left( \frac{\rho_a}{0.3 {\rm GeV} {\rm cm}^{-3}} \right)
\left( \frac{10^{-3}}{\beta_a} \right)^3
\left( \frac{10^{-5} {\rm eV}}{m_a} \right)^4 ,
\label{eqn:nbara}
\end{equation}
where the enegy density of the cosmic axions $ \rho_a $ 
is taken to be equal to that of the galactic dark halo
$ \rho_{\rm halo} \simeq 0.3 {\rm GeV} {\rm cm}^{-3} $.
The dissipation of the coherent axions is characterized
by their energy spread as
\begin{equation}
\gamma_a \sim \beta_a^2 m_a / \hbar
= 0.02 \gamma
\left( \frac{\beta_a^2 Q }{10^{-6} \times 2 \times 10^4} \right) ,
\end{equation}
where the resonant condition $ m_a \simeq \hbar \omega_c $ is considered.
All the atoms are prepared in the lower state so that
\begin{equation}
{\bar n}_b = 0 .
\label{eqn:nbarb}
\end{equation}
The atomic dissipation is determined by its lifetime $ \tau_b $
as $ \gamma_b = \tau_b^{-1} $.
The lifetime of the Rydberg state is typically
$ \tau_b \sim 10^{-3} {\rm s} $ for $ n \sim 100 $ in the vacuum
\cite{Gallagher94}.
Since the transitions to the off-resonant states are highly suppressed
in the resonant cavity, the atomic lifetime may even be longer.
Hence, the damping rate of Rydberg atoms is estimated to be at most
\begin{equation}
\gamma_b = 6.6 \times 10^{-13} {\rm eV} \hbar^{-1}
\left( \frac{10^{-3} {\rm s}}{\tau_b} \right) ,
\end{equation}
which is actually much smaller than the photon damping rate $ \gamma $.

\section{Interactions in the resonant cavity}
\label{sec:interactions}

In this section, we provide the interaction Hamiltonians
in terms of the quantum oscillators.
They determine the quantum evolution of the axion-photon-atom system
in the resonant cavity,
which will be investigated in detail in the following sections.

\subsection{Axion-photon interaction}

The axion-photon interaction under a strong static magnetic field
with flux density $ {\mbox{\boldmath $ B $}}_0 $
is described by the Lagrangian density
\begin{equation}
{\cal L}_a = \hbar^{1/2} \epsilon_0 g_{a \gamma \gamma}
\phi {\mbox{\boldmath $ E $}} \cdot {\mbox{\boldmath $ B $}}_0 ,
\label{eqn:La}
\end{equation}
where $ \hbar^{1/2} $ and $ \epsilon_0 $ (dielectric constant)
are explicitly factored out
so that the Lagrangian density has the right dimension
$ {\cal L}_a \sim \hbar {\rm s}^{-1} {\rm m}^{-3}
\sim {\rm eV} {\rm m}^{-3} $
with the axion-photon-photon coupling constant
$ g_{a \gamma \gamma} \sim {\rm eV}^{-1} $, as given below.
The axion-photon-photon coupling constant is calculated
\cite{WWBT78,KS85} as
\begin{equation}
g_{a \gamma \gamma} = c_{a \gamma \gamma}
\frac{\alpha}{2 \pi^2} \frac{m_a}{f_\pi m_\pi} \frac{(1 + Z)}{{\sqrt Z}} ,
\label{eqn:gagg}
\end{equation}
where $ Z = m_u / m_d $, and
\begin{equation}
c_{a \gamma \gamma} = \frac{E}{C} - \frac{2(4 + Z)}{3(1 + Z)}
\end{equation}
with
\begin{equation}
E = {\rm Tr} Q_{\rm PQ} Q_{\rm em}^2 , \
C \delta_{ab} = {\rm Tr} Q_{\rm PQ} \lambda_a \lambda_b .
\end{equation}
The parameter $ c_{a \gamma \gamma} $ represents
the variation of the axion-photon-photon coupling
depending on the respective Peccei-Quinn models such as
the so-called KSVZ \cite{KSVZ}
and DFSZ \cite{DFSZ}.

The original Lagrangian density for the axion-photon-photon coupling
in Eq. (\ref{eqn:La}) provides the effective interaction Hamiltonian
between the coherent axion mode $ a $
and the resonant radiation mode $ c $,
\begin{equation}
H_{ac} = \hbar \kappa ( a^\dagger c + a c^\dagger ) .
\label{eqn:H-ac}
\end{equation}
The axion-photon conversion in the cavity $ {\cal V}_1 $
is well described with this interaction Hamiltonian.
The coupling constant $ \kappa $ is determined
for $ \omega_c \simeq m_a / \hbar $
by considering the relations
$ | \langle \eta_a | \phi^\pm | \eta_a \rangle | \simeq
\hbar ( \rho_a / m_a^2 )^{1/2} $ from Eq. (\ref{eqn:rhoa-na})
and
$ | \langle \eta_a | a | \eta_a \rangle | = {\bar n}_a^{1/2} $
from Eq. (\ref{eqn:na})
in calculating
$ \langle \eta_a | \int_{{\cal V}_1} ( - {\cal L}_a ) d^3 x | \eta_a \rangle
\simeq \langle \eta_a | H_{ac} | \eta_a \rangle $
\cite{OMY96}:
\begin{eqnarray}
\kappa & = & \hbar^{1/2} g_{a \gamma \gamma} \epsilon_0^{1/2}
B_{\rm eff} \left[ \left( \frac{\beta_a m_a}{2 \pi \hbar} \right)^3
\frac{V_1}{2} \right]^{1/2}
\nonumber \\ 
& = & 4 \times 10^{-26}{\rm eV} \hbar^{-1}
\left( \frac{g_{a \gamma \gamma}}{1.4 \times 10^{-15}{\rm GeV}^{-1}} \right)
\left( \frac{B_{\rm eff}}{4{\rm T}} \right)
\nonumber \\
& \times & \left( \frac{\beta_a m_a }{10^{-3}
\times 10^{-5}{\rm eV}} \right)^{3/2}
\left( \frac{V_1}{5000{\rm cm}^3} \right)^{1/2} ,
\label{eqn:kappa}
\end{eqnarray}
where
\begin{equation}
B_{\rm eff} = \zeta_1 G B_0 ,
\end{equation}
and $ B_0 $ is the maximal density of the external magnetic flux.
The axion-photon-photon coupling constant $ g_{a \gamma \gamma} $
is taken here to be the value expected from the DFSZ axion model
\cite{DFSZ} at $ m_a = 10^{-5} {\rm eV} $.

This effective axion-photon coupling $ \kappa $ apparently
involves the $ \beta_a $ dependence coming from the normalization
given in Eq. (\ref{eqn:sigma-a}).
This $ \beta_a $ dependence is, however, canceled
with that of $ {\bar n}_a $ given in Eq. (\ref{eqn:na})
in calculating the signal rate which is actually proportional
to $ ( \kappa / \gamma )^2 {\bar n}_a $.
The form factor for the magnetic field is given by
\begin{equation}
G = \zeta_1^{-1} V_1^{-1/2} \left| \int_{{\cal V}_1} d^3 x
{\mbox{\boldmath $ \alpha $}}_1 ({\bf x}) \cdot
[ {\mbox{\boldmath $ B $}}_0 ({\bf x})/B_0 ] \right|
\label{eqn:G}
\end{equation}
with
\begin{equation}
\zeta_1 = \left[ \int_{{\cal V}_1} d^3 x
| {\mbox{\boldmath $ \alpha $}}_1 ({\bf x}) |^2 \right]^{1/2} .
\label{eqn:zeta1}
\end{equation}
This additional factor $ \zeta_1 $ ($ < 1 $
as seen from Eq.(\ref{eqn:moden}))
represents the effective reduction of the axion-photon conversion
which is due to the fact that the magnetic field
is applied only in the conversion cavity $ {\cal V}_1 $.
We may obtain the effective magnetic field strength
$ B_{\rm eff} \simeq 4 {\rm T} $,
as taken in Eq.(\ref{eqn:kappa}),
by using typically a magnet of $ B_0 \simeq 7{\rm T} $
and the cavity system with $ G = {\sqrt{0.7}} $
of $ {\rm TM}_{010} $ mode and $ \zeta_1 \simeq 0.7 $
for the conversion cavity.

It should here be remarked that the possible interaction term
$ \hbar \kappa ( a c + a^\dagger c^\dagger ) $
is not included in Eq. (\ref{eqn:H-ac}).  This is justified as follows.
The characteristic time scale for the evolution of the system
considered in the present scheme is placed by the lifetime of photons
$ \tau_\gamma = \gamma^{-1} $ in the cavity.
Hence, although the interaction term
$ \hbar \kappa ( a c + a^\dagger c^\dagger ) $
representing the processes with energy change of $ \hbar \omega_c + m_a $
is also obtained from the original axion-photon-photon coupling
$ {\cal L}_a $, its effects are suppressed sufficiently for $ Q \gg 1 $
by the energy conservation in this time scale $ \tau_\gamma $.

\subsection{Atom-photon interaction}

We next consider the interaction
between the Rydberg atoms and the resonant photons.
The Rydberg atoms are utilized for counting photons in the cavity.
They are excited by absorbing the photons
through the electric dipole transition with frequency $ \omega_b $.
The emission and absorption of photon by the two-level atomic system
is described by the interaction Hamiltonian
\begin{eqnarray}
H_\Omega &=& \hbar \Omega [ D^+ c + D^- c^\dagger ]
\nonumber \\
& \simeq & \hbar \Omega [ b^\dagger c + b c^\dagger ] .
\label{eqn:H-Omega}
\end{eqnarray}
Here, we may assume for simplicity, though not essential,
that the mode vector field is described
in terms of a uniform complex polarization vector
$ {\mbox{\boldmath $ \epsilon $}}_2 $
($ | {\mbox{\boldmath $ \epsilon $}}_2 | = 1 $)
and a real profile function $ f_2 ({\bf x}) $ as
\begin{equation}
{\mbox{\boldmath $ \alpha $}}_2 ({\bf x})
= {\mbox{\boldmath $ \epsilon $}}_2 {\bar V}_2^{-1/2} f_2 ({\bf x})
\label{eqn:aef}
\end{equation}
with $ {\bar V}_2 \equiv
| {\mbox{\boldmath $ \alpha $}}_2 ({\bf x}) |_{\rm max}^{-2} $.
The profile function is normalized for convenience
by the condition $ | f_2 ({\bf x}) |_{\rm max} = 1 $
at the antinodal positions.
Then, the intrinsic atom-photon coupling constant $ \Omega $
is evaluated at the antinodal position
in terms of the electric dipole transition matrix element
$ d = | {\mbox{\boldmath $ \epsilon $}}_2 \cdot {\mbox{\boldmath $ d $}} | $
projected to the direction of polarization:
\begin{equation}
\Omega = \frac{d}{\hbar}
\left( \frac{\hbar \omega_c}{2 \epsilon_0 {\bar V}_2} \right)^{1/2} ,
\label{eqn:Omega}
\end{equation}
where $ {\bar V}_2 \sim V_2 $
is expected in accordance with the electric field
normalization condition (\ref{eqn:moden}).
The atom-photon coupling at an arbitrary atomic position in the cavity
is also given with the profile function by
\begin{equation}
\Omega ({\bf x}) = \Omega f_2 ({\bf x}) .
\label{eqn:Omegax}
\end{equation}

In the actual experimental system, the Rydberg atoms are injected
as a uniform beam.  Then, certain number $ N $ of Rydberg atoms
are constantly present in the cavity.
Such an ensemble of identical atoms behaves as a collective system
in the interaction with the resonant radiation mode.
The novelty of Rydberg atom physics is to provide situations
where this collective behavior appears for a relatively small number
of atoms, typically $ N \sim 10^3 - 10^6 $
\cite{Haroche85}.
Suppose for simplicity that the $ N $ atoms are at the antinodal position
($ \Omega ({\bf x}) = \Omega $) in the detection cavity.
Then, the effective coupling between the radiation mode and
the $ N $ atoms is actually given by
\begin{equation}
H_{bc} = \sum_{i=1}^N H_{\Omega} ( b_i , c )
= \hbar \Omega_N ( b^\dagger c + b c^\dagger ) ,
\label{eqn:H-bc}
\end{equation}
where the collective atomic mode operator is defined by
\begin{equation}
b \equiv \frac{1}{\sqrt N} \sum_{i=1}^N b_i
\label{eqn:b}
\end{equation}
satisfying the commutation relation $ [ b , b^\dagger ] = 1 $.
(Hereafter we may use without confusion
the operators $ b $ and $ b^\dagger $ for the collective atomic mode
rather than the single atomic mode.)
The collective atom-photon coupling $ \Omega_N = \Omega {\sqrt N} $
\cite{Haroche85}
is estimated typically as
\begin{equation}
\Omega_N = 1 \times 10^{-10}{\rm eV} \hbar^{-1}
\left( \frac{\Omega}{5 \times 10^3{\rm s}^{-1}} \right)
\left( \frac{N}{10^3} \right)^{1/2} .
\label{eqn:OmegaN}
\end{equation}
Hence, if the number of Rydberg atoms is as large as $ N \sim 10^3 $,
this collective coupling can be comparable to the cavity damping rate
$ \gamma $ as given in Eq. (\ref{eqn:gamma}).
When the atomic motion and distribution in the cavity are taken into account,
the atom-photon coupling should be modified suitably.
This point will be treated in Sec. \ref{sec:sensitivity}.

\section{Quantum evolution of the system}
\label{sec:evolution}

We here formulate the quantum dynamics of interacting oscillators
with dissipation.
A systematic procedure is developed in the Liouville picture
to calculate the quantum averages of the particle numbers
and higher-order correlations.
In practice, a series of master equations is derived
for such quantities
from the Fokker-Planck equation based on the Liouville picture.
(The quantum evolution of the system can also be described
in the Langevin picture, which is considered
in Appendix \ref{appendix:langevin}.)
By applying this formulation, the quatum evolution of
the axion-photon-atom system in the resonant cavity is determined
with the effective interaction Hamiltonians
presented in the preceeding section.
Then, we can calculate, in particular, the number of the atoms
in the upper state which are excited by absorbing the axion-converted
and thermal photons.

\subsection{Evolution in Liouville picture}

The reduced density matrix of the damped oscillators $ q_i $
(such as $ a , b , c $) interacting each other obeys the Liouville equation
\cite{Louisell90},
\begin{equation}
\frac{d \rho}{dt} = \frac{1}{i \hbar} [ H , \rho ]
+ \Lambda \rho .
\label{eqn:rho-eqn}
\end{equation}
The Liouvillian relaxations are represented by the operator
$ \Lambda \rho $, which may explicitly be given by
\begin{eqnarray}
\Lambda \rho &=& \sum_i 
\frac{\gamma_i}{2} \left[ 2 q_i \rho q_i^\dagger
- q_i^\dagger q_i \rho - \rho q_i^\dagger q_i \right]
\nonumber \\
&+& \sum_i \gamma_i {\bar n}_i
\left[ q_i^\dagger \rho q_i + q_i \rho q_i^\dagger
- q_i^\dagger q_i \rho - \rho q_i q_i^\dagger \right] ,
\label{eqn:Lam-rho}
\end{eqnarray}
where $ \gamma_i $ and $ {\bar n}_i $ are the damping rates
and equilibrium occupation numbers, respectively.
The total Hamiltonian is given by
\begin{equation}
H = \sum_i \hbar \omega_i q_i^\dagger q_i
+ \sum_{i \not= j} \hbar \Omega_{ij} (t) q_i^\dagger q_j ,
\label{eqn:H}
\end{equation}
where $ \Omega_{ij} (t) = \Omega^*_{ji} (t) $
represent the interaction terms, which may be time-dependent,
as is the case for the interaction between photons and moving atoms
in the cavity (see Eq. (\ref{eqn:Hmatrix})).

The solution of the Liouville equation is generally obtained
in terms of the creation and annihilation operators as
\begin{equation}
\rho = \rho ( q_i , q_i^\dagger , t ) .
\label{eqn:rho-sol}
\end{equation}
The quantum average of a physical quantity represented by
a relevant operator $ {\cal O} $ is evaluated with
this density matrix by the formula
\begin{equation}
{\bar {\cal O}} (t) = \langle {\cal O} \rangle
\equiv {\rm Tr}[ {\cal O} ( q_i , q_i^\dagger , t )
\rho ( q_i , q_i^\dagger , t ) ] .
\label{eqn:O-bar}
\end{equation}
In practical calculations, it may be useful to take the coherent state basis
\cite{Louisell90}.
Then, the density operator $ \rho $ is represented
by the ``classical" time-dependent distribution function
$ P( \alpha_i , \alpha_i^* , t ) $,
and the quantum average is calculated by
\begin{equation}
{\bar {\cal O}} (t) = \int \prod_i d^2 \alpha_i
{\cal O}_{\rm n} ( \alpha_i , \alpha_i^* , t )
P( \alpha_i , \alpha_i^* , t ) .
\end{equation}
Here the subscript ``n" means the the expression of the operator
$ {\cal O} $ when it is written in the normal ordered form.
For example, $ {\cal O}_{\rm n} ( \alpha , \alpha^* )
= \alpha^* \alpha + 1 $ for $ {\cal O} ( q , q^\dagger ) = q q^\dagger $.

The Liouville equation is expressed in the coherent state basis
as the Fokker-Planck equation to determine the distribution function $ P $:
\begin{eqnarray}
\frac{\partial P}{\partial t}
&=& i {\cal H}_{ij} (t) \frac{\partial}{\partial \alpha_i} ( \alpha_j P )
- i {\cal H}^*_{ij} (t) \frac{\partial}{\partial \alpha_i^*} ( \alpha_j^* P )
\nonumber \\
&+& {\cal D}_{ij}
\frac{\partial^2 P}{\partial \alpha_i \partial \alpha_j^*} .
\label{eqn:F-Peq}
\end{eqnarray}
Here, the effective Hamiltonian and diffusion term
for the damped oscillators are represented by the matrices
\begin{eqnarray}
{\cal H}_{ij} (t) &=& \left( \omega_i - \frac{i}{2} \gamma_i \right)
\delta_{ij}
+ \Omega_{ij} (t) ( 1 - \delta_{ij} ) ,
\label{eqn:Hij(t)} \\
{\cal D}_{ij} &=& \gamma_i {\bar n}_i \delta_{ij} .
\label{eqn:Dij}
\end{eqnarray}
They are given explicitly for the axion-photon-atom system
under consideration by
\begin{equation}
{\cal H} (t) = \left( \begin{array}{ccc}
\omega_b - \frac{i}{2} \gamma_b & \Omega (t) & 0 \\
\Omega (t) & \omega_c - \frac{i}{2} \gamma_c & \kappa \\
0 & \kappa & \omega_a - \frac{i}{2} \gamma_a \\
\end{array} \right)
\label{eqn:Hmatrix}
\end{equation}
with $ \omega_a = m_a / \hbar $, and
\begin{equation}
{\cal D} = \left( \begin{array}{ccc}
\gamma_b {\bar n}_b & 0 & 0 \\
0 & \gamma_c {\bar n}_c & 0 \\
0 & 0 & \gamma_a {\bar n}_a \end{array} \right) .
\end{equation}
The collective atom-photon coupling becomes time-dependent
through the atomic motion along the $ x $ axis with velocity $ v $ as
\begin{equation}
\Omega (t) = \Omega_N f(vt) .
\label{eqn:Omt}
\end{equation}
Here, it is assumed for simplicity that the $ N $ atoms are injected
together as a pulsed beam.
This time-dependence is determined by the atomic velocity $ v $
and the electric field profile $ f(x) \equiv f_2 (x,y_0,z_0) $
of the detection cavity along the atomic beam
which is injected in the $ x $ direction
from the point $ (0,y_0,z_0) $ in the $ y $-$ z $ plane.

\subsection{Master equations}

We may be interested in the self-adjoint multiple moments
of the oscillators,
\begin{eqnarray}
& & {\cal N}^{(2p)}_{i_1 \cdots i_p j_1 \cdots j_p} (t)
\equiv \langle q_{i_1}^\dagger \cdots q_{i_p}^\dagger
q_{j_1} \cdots q_{j_p} \rangle
\nonumber \\
& & = \int \prod_i d^2 \alpha_i \alpha_{i_1}^* \cdots \alpha_{i_p}^*
\alpha_{j_1} \cdots \alpha_{j_p} P( \alpha_i , \alpha_i^* , t ) ,
\end{eqnarray}
where $ 2p $ denotes the number of involved operators.
A series of master equations is then obtained
for these moments by considering the Fokker-Planck equation
for the distribution function:
\begin{equation}
\frac{d {\cal N}^{(2p)}}{dt} = - i {\cal H}^{(2p)} (t) {\cal N}^{(2p)}
+ {\cal D}^{(2p-2)} {\cal N}^{(2p-2)} .
\end{equation}
Here the linear operators
$ {\cal H}^{(2p)} $ and $ {\cal D}^{(2p-2)} $
acting on the moments are defined by
\begin{eqnarray}
( {\cal H}^{(2p)} (t) {\cal N}^{(2p)} )_{IJ}
& \equiv & \sum_k
{\cal N}^{(2p)}_{i_1 \cdots i_p j_1 \cdots j^\prime_k \cdots j_p}
{\cal H}_{j^\prime_k j_k}^{\rm T} (t)
\nonumber \\
&-& \sum_k {\cal H}_{i_k i^\prime_k}^* (t)
{\cal N}^{(2p)}_{i_1 \cdots i^\prime_k \cdots i_p j_1 \cdots j_p} , \\
( {\cal D}^{(2p-2)} {\cal N}^{(2p-2)} )_{IJ}
& \equiv & \sum_{k,l}
{\cal D}_{i_k j_l} {\cal N}^{(2p-2)}_{I^\prime_k J^\prime_l}
\end{eqnarray}
with abbreviation of the indices
$ I \equiv i_1 \cdots i_p $, $ J \equiv j_1 \cdots j_p $,
$ I^\prime_k \equiv i_1 \cdots i_{k-1} i_{k+1} \cdots i_p $
and $ J^\prime_l \equiv j_1 \cdots j_{l-1} j_{l+1} \cdots j_p $.

We examine, in particular, the second-order moment
$ {\cal N}_{ij} (t) \equiv {\cal N}^{(2)}_{ij} (t) $.
The number of the atoms which are excited by absorbing the photons
is given by the diagonal component of $ {\cal N} (t) $ as
\begin{equation}
n_b (t) = \langle b^\dagger b \rangle = {\cal N}_{bb} (t) .
\end{equation}
The master equation for $ {\cal N} (t) $ reads
\begin{equation}
\frac{d{\cal N}}{dt} = -i {\cal N} {\cal H}^{\rm T} (t)
+ i {\cal H}^*(t) {\cal N} + {\cal D} .
\label{eqn:Nmaster}
\end{equation}
This linear equation with an inhomogeneous term may be solved formally
as follows.  We first introduce a new matrix $ {\cal N}^\prime (t) $
instead of $ {\cal N} (t) $ by
\begin{equation}
{\cal N}^\prime (t)
= {\cal U}^{-1 *} (t) {\cal N} (t) {\cal U}^{-1 {\rm T}} (t) .
\label{eqn:Nprmt}
\end{equation}
The linear transformation $ {\cal U}(t) $ representing the time evolution
due to $ {\cal H} (t) $ is given by
\begin{equation}
{\cal U}(t)
= {\rm P} \left[ -i \exp \int_0^t {\cal H} ( \tau ) d \tau \right]
\label{eqn:Ut}
\end{equation}
satisfying the equation $ d{\cal U}/dt = -i {\cal H} (t) {\cal U} $
with $ {\cal U} (0) = {\bf 1} $,
where ``P" denotes the chronological product.
Then, the master equation (\ref{eqn:Nmaster})
is reduced as
\begin{equation}
\frac{d{\cal N}^\prime}{dt}
= {\cal U}^{-1 *} (t) {\cal D} {\cal U}^{-1 {\rm T}} (t) .
\end{equation}
By considering the initial condition $ {\cal N}^\prime (0) = {\cal N}(0) $
with $ {\cal U}(0) = {\bf 1} $ and Eq. (\ref{eqn:Dij}) for $ {\cal D} $,
we can solve the above equation as
\begin{equation}
{\cal N}^\prime_{ij} (t)
= {\cal B}^k_{ij} (t) {\bar n}_k + {\cal N}_{ij} (0) ,
\end{equation}
where
\begin{equation}
{\cal B}^k_{ij} (t) = \int_0^t \gamma_k
{\cal U}^{-1 *}_{ik} ( \tau ) {\cal U}^{-1 {\rm T}}_{kj} ( \tau ) d \tau .
\end{equation}
Then, we obtain the solution as
\begin{equation}
{\cal N}_{ij} (t) = {\cal R}_{ij}^k (t) {\bar n}_k
+ {\cal U}^*_{ik} (t) {\cal N}_{kl} (0) {\cal U}^{\rm T}_{lj} (t) ,
\label{eqn:Ntsol}
\end{equation}
where
\begin{equation}
{\cal R}^k_{ij} (t)
= \left[ {\cal U}^* (t) {\cal B}^k (t) {\cal U}^{\rm T} (t) \right]_{ij} .
\end{equation}
The same result is also obtained in the Langevin picture
(see Appendix \ref{appendix:langevin}).
This solution approaches to certain asymptotic value
after a long enough time as
\begin{equation}
{\cal N}_{ij} (t) \approx {\cal R}_{ij}^k (t) {\bar n}_k \
( t \gg {\rm max} [ \gamma_i^{-1} ] ) .
\end{equation}
In practical calculations, the master equation (\ref{eqn:Nmaster})
will be solved numerically with the time-dependent
effective Hamiltonian $ {\cal H} (t) $
as given in Eq. (\ref{eqn:Hmatrix}).

\section{Aspects of axion-photon-atom interaction}
\label{sec:aspects}

We can see some characteristic properties of the axion-photon-atom
interaction in the resonant cavity by examining the simple case
with the constant atom-photon coupling
$ \Omega (t) = \Omega_N $ for the effective Hamiltonian
$ {\cal H}(t) = {\cal H} $ in Eq. (\ref{eqn:Hmatrix}).
In this case, as derived in the Appendix \ref{appendix:analytic},
the analytic solution is obtained for the particle numbers
(with the condition $ {\bar n}_b = 0 $) as
\begin{equation}
n_i (t) = \langle q_i^\dagger q_i \rangle
= r_{ic} (t) {\bar n}_c + r_{ia} (t) {\bar n}_a ,
\end{equation}
where
\begin{equation}
r_{ij}(t) = \sum_{m,n} g_{ij}^{m*} g_{ij}^n
\left[ \left( 1 - \frac{\gamma_j}{\Lambda_{mn}} \right)
{\rm e}^{- \Lambda_{mn} t}
+ \frac{\gamma_j}{\Lambda_{mn}} \right]
\end{equation}
with
\begin{eqnarray}
g^k_{ij} &=& \lim_{s \rightarrow - i \lambda_k}
( s + i \lambda_k ) ( s {\bf 1} + i {\cal H} )_{ij}^{-1} ,
\label{eqn:gkij} \\
\Lambda_{mn} &=& - i ( \lambda_m^* - \lambda_n ).
\end{eqnarray}
Here, the atomic damping rate $ \gamma_b $ may be neglected
for simplicity, since it is sufficiently smaller
than $ \gamma_a $ and $ \Omega_N $
($ \gamma_b \sim 0.001 \gamma $ with $ \tau_b \sim 10^{-3} {\rm s} $
for $ \omega_c \sim 10^{-5} {\rm eV} $ and $ Q \sim 10^4 $).
The condition $ \omega_b = \omega_c $ may also be taken for definiteness,
since the atomic transition frequency should be tuned
almost equal to the cavity frequency.
Then, the eigenvalues of the Hamiltonian $ {\cal H} $ are given by
\begin{eqnarray}
\lambda_1 &=& \omega_c - \frac{i}{4} \gamma
+ i \frac{( \gamma^2 - 16 \Omega_N^2 )^{1/2}}{4} ,
\label{eqn:lm1} \\
\lambda_2 &=& \omega_c - \frac{i}{4} \gamma
- i \frac{( \gamma^2 - 16 \Omega_N^2 )^{1/2}}{4} ,
\label{eqn:lm2} \\
\lambda_3 &=& \omega_a - \frac{i}{2} \gamma_a ,
\label{eqn:lm3}
\end{eqnarray}
where
\begin{equation}
( \gamma^2 - 16 \Omega_N^2 )^{1/2}
= \left\{ \begin{array}{ll}
{\sqrt{\gamma^2 - 16 \Omega_N^2}} & ( \Omega_N / \gamma \leq 1/4 ) \\
i {\sqrt{16 \Omega_N^2 - \gamma^2}} & ( \Omega_N / \gamma > 1/4 )
\end{array} \right. .
\end{equation}
If the number of atoms $ N $ (or the atomic beam intensity $ I_{\rm Ryd} $)
is not so large giving $ \Omega_N / \gamma < 1/4 $, the damping rate
of the eigenmode of $ \lambda_1 $ is smaller than $ \gamma / 2 $,
and that of $ \lambda_2 $ lies between $ \gamma / 2 $ and $ \gamma $.
On the other hand, in the strong coupling region
of $ \Omega_N / \gamma > 1/4 $
for the collective atom-photon interaction
the eignemodes of $ \lambda_1 $ and $ \lambda_2 $
form a doublet around the frequency $ \omega_c $
with the same damping rate $ \gamma / 2 $ (Rabi splitting).

Among the damping rates $ {\rm Re} [ \Lambda_{mn} ] $
of the respective terms in the factors $ r_{ij} (t) $
representing the contributions of thermal photons and axions,
$ {\rm Re} [ \Lambda_{33} ] = \gamma_a $
($ \sim 0.01 \gamma $ for $ \beta_a \sim 10^{-3} $
and $ Q \sim 10^4 $, typically) is the smallest one.
The rate
$ {\rm Re} [ \Lambda_{11} ] \simeq 4 \gamma ( \Omega_N / \gamma )^2 $
for $ \Omega_N / \gamma \sim 0.1 $ may also be comparable
to the smallest rate $ {\rm Re} [ \Lambda_{33} ] $.
The atomic transit time through the cavity is, on the other hand,
given by
\begin{equation}
t_{\rm tr} = L / v
\end{equation}
with the detection cavity length $ L $ and the atomic velocity $ v $.
This transit time provides the effective cut-off
for the axion-photon-atom interaction in the cavity.
(Here we assume for simplicity that the atoms have the uniform velocity.)
It is typically $ t_{\rm tr} \simeq 400 \tau_\gamma $
with $ L = 0.2 {\rm m} $ and $ v = 350 {\rm m} {\rm s}^{-1} $
for $ m_a = 10^{-5} {\rm eV} $ and $ Q = 2 \times 10^4 $
in the case of the detection apparatus such as CARRACK I.
Hence, the transit time can be regareded to be long enough compared
to $ ( {\rm Re} [ \Lambda_{mn} ] )^{-1} $, i.e.,
\begin{equation}
t_{\rm tr} > {\rm several} \ \gamma_a^{-1} ,
\end{equation}
so that the respective particle numbers
will almost reach the asymptotic values as
\begin{equation}
r_{ij}(t_{\rm tr}) \approx r_{ij} ( \infty )
= \sum_{m,n} g_{ij}^{m*} g_{ij}^n \frac{\gamma_j}{\Lambda_{mn}} .
\label{eqn:rij-inf}
\end{equation}

By using Eqs. (\ref{eqn:lm1}) -- (\ref{eqn:lm3})
and the explicit matrix form (\ref {eqn:siH})
for $ ( s {\bf 1} + i {\cal H} )^{-1} $
given in Appendix \ref{appendix:analytic},
we can calculate the coefficients $ g_{ij}^k $ in Eq. (\ref{eqn:gkij}).
Then, we can show the relations
\begin{equation}
r_{bc} ( \infty ) = r_{cc} ( \infty ) = 1 .
\end{equation}
This implies that if the axion-photon interaction is turned off,
the numbers of the photons and excited atoms reach
the same asymptotic value $ {\bar n}_c $
of the thermal photon number at $ T_c $
\cite{Haroche85}:
\begin{equation}
n_b [ c \rightarrow b ] \approx n_c [ c \rightarrow c ]
\approx {\bar n}_c .
\end{equation}
The number of axion-converted photons is, on the other hand,
given approximately by
\begin{equation}
n_c [ a \rightarrow c ] \approx r_{ca} ( \infty ) {\bar n}_a .
\end{equation}
The number of excited atoms due to the axion-converted photons
is also given by
\begin{equation}
n_b [ a \rightarrow c \rightarrow b ]
\approx r_{ba} ( \infty ) {\bar n}_a .
\end{equation}
These contributions from the axions are essentially determined
by the factors
\begin{equation}
g^m_{ca} , \ g^m_{ba} \ \propto \
\frac{\kappa}{( \lambda_m - \lambda_k )( \lambda_m - \lambda_l )} ,
\label{eqn:gmia}
\end{equation}
where $ m \not= k , l $ and $ k \not = l $.
The detuning of axion mass from the cavity frequency is given by
\begin{equation}
\Delta \omega_a \equiv \omega_a - \omega_c .
\end{equation}
Then, the above factors are enhanced if the axion detuning
$ \Delta \omega_a $ lies in the range where the resonant conditions
$ \lambda_3 \simeq \lambda_1 $ and/or $ \lambda_3 \simeq \lambda_2 $
are satisfied.

\begin{figure}
\begin{center}
\includegraphics[height=14cm]{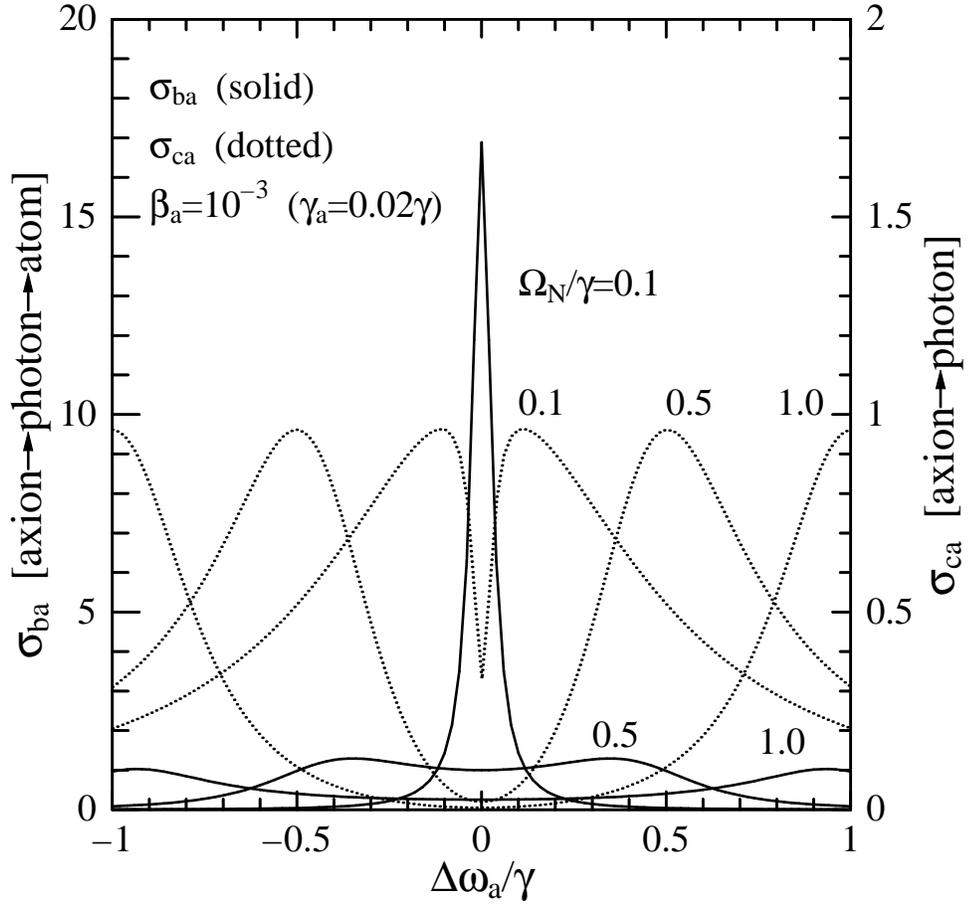}
\caption{The form factors for the axion conversion versus the axion detuning
for the case with constant atom-photon coupling.
The axion width is taken to be $ \gamma_a = 0.02 \gamma $
with $ \beta_a = 10^{-3} $ and $ Q = 2 \times 10^4 $.
\label{fig:sgiada1}}
\end{center}
\end{figure}

In the leading order of the axion-photon coupling $ \kappa $
with Eq. (\ref{eqn:gmia}),
the factors for the axion contributions are given with
$ r_{ca} ( \infty ) , \ r_{ba} ( \infty ) \propto ( \kappa / \gamma )^2 $.
Then, by noting the relation
$ ( \kappa / \gamma)^2 {\bar n}_a \propto ( \rho_a / m_a ) V_1 $
from Eqs. (\ref{eqn:na}) and (\ref{eqn:kappa})
it is found that the axion-converted photons $ n_c [ a \rightarrow c ] $
and the number of atoms $ n_b [ a \rightarrow c \rightarrow b ] $
excited by such photons are both proportional to the number of axions
contained in the conversion cavity.
It is here relevant to define the form factors for the axion contributions
with respect to the axion detuning
($ t_{\rm tr} > {\rm several} \ \gamma_a^{-1} \gg \gamma^{-1} $)
by
\begin{equation}
\sigma_{ia} ( \Delta \omega_a )
= \frac{r_{ia} ( t_{\rm tr} )}{[ \kappa / ( \gamma / 2 ) ]^2}
\simeq \frac{r_{ia} ( \infty )}{[ \kappa / ( \gamma / 2 ) ]^2} \ \
( i = b , c ) .
\end{equation}
These form factors $ \sigma_{ba} ( \Delta \omega_a ) $
for $ a \rightarrow c \rightarrow b $ (solid lines)
and $ \sigma_{ca} ( \Delta \omega_a ) $
for $ a \rightarrow c $ (dotted lines)
are plotted together in Fig. \ref{fig:sgiada1}
for some typical values of the atom-photon coupling,
$ \Omega_N / \gamma = 0.1 , 0.5 , 1.0 $.

The behavior of $ \sigma_{ba} ( \Delta \omega_a ) $ is, in particular,
understood by noting its dominant contribution which is in fact given
by the term of $ | g_{ba}^{m=3} |^2 $ ($ \gamma_a \ll \gamma $) as
\begin{equation}
\sigma_{ba} ( \Delta \omega_a )
\simeq \frac{4 \Omega_N^2 \gamma^2}
{4 \Delta \omega_a^2 \gamma ( \gamma - 4 \gamma_a )
+ ( 4 \Delta \omega_a^2 - 4 \Omega_N^2 + \gamma_a \gamma )^2} .
\label{eqn:sgba0}
\end{equation}
The peak of this form factor appears at
\begin{equation}
\Delta \omega_a ({\rm peak}) = \left\{ \begin{array}{ll}
0 & ( \Omega_N \leq {\bar \Omega} ) \\
\pm {\sqrt{\Omega_N^2 - {\bar \Omega}^2}}
& ( \Omega_N > {\bar \Omega} )
\end{array} \right. ,
\end{equation}
where $ {\bar \Omega} = \gamma /{\sqrt 8} + O( \gamma_a ) $.
(Although $ \sigma_{ba} (0) $ is apparently divergent
for $ \Omega_N = \frac{1}{2} ( \gamma_a \gamma )^{1/2} $
in the approximate formula of Eq. (\ref{eqn:sgba0}),
the other contributions also become significant
around $ \Delta \omega_a = 0 $
so as to give a finite value of $ \sigma_{ba} (0) $.)
It is found that for $ \Omega_N \gtrsim \gamma / 4 $
the peak value of the axion signal decreases due to the Rabi splitting,
which approaches to $ \sigma_{ba} ( \Delta \omega_a ({\rm peak}) ) \simeq 1 $
for $ ( \Omega_N / \gamma )^2 \gg 1 $.
On the other hand, for the appropriate atom-photon coupling such as
\begin{equation}
\Omega_N \sim ( \gamma_a \gamma )^{1/2}
\label{eqn:OmN1}
\end{equation}
the signal is enhanced significantly, as observed in Fig. \ref{fig:sgiada1}
($ \Omega_N = 0.1 \gamma $ and $ \gamma_a = 0.02 \gamma $),
by virtue of the narrow width (long coherence)
of the galactic axions as
\begin{equation}
\sigma_{ba} ( \Delta \omega_a ) \sim ( \gamma_a / \gamma )^{-1} ,
\label{eqn:sgba1}
\end{equation}
when the axion detuning becomes small enough to satisfy the condition
\begin{equation}
| \Delta \omega_a | \lesssim \gamma_a .
\label{eqn:Doma1}
\end{equation}
Here, the energy uncertainty $ \sim \hbar / t_{\rm tr} $
due to the atomic motion is assumed to be smaller
than the axion width $ \hbar \gamma_a $.

The form factor $ \sigma_{ca} ( \Delta \omega_a ) $
determines the equilibrium number
$ r_{ca} ( \infty ) {\bar n}_a
= 4 \sigma_{ca} ( \Delta \omega_a )( \kappa / \gamma )^2 {\bar n}_a $
of the axion-converted photons in the cavity.
As seen in Fig. \ref{fig:sgiada1},
its peak value is $ \sigma_{ca} \simeq 1 $
almost independently of the atom-photon coupling $ \Omega_N $.
It is here, in particular, interesting to observe a narrow dip
in $ \sigma_{ca} ( \Delta \omega_a ) $
around $ \Delta \omega_a = 0 $ for $ \Omega_N / \gamma = 0.1 $.
This indicates that the axion-converted photons are efficiently
absorbed by the atoms
for $ \Omega_N / \gamma \sim ( \gamma_a / \gamma )^{1/2} $.
For larger $ \Omega_N $, two separate peaks appear
in $ \sigma_{ca} ( \Delta \omega_a ) $ due to the Rabi splitting.

Some characteristic features concerning the axion-photon-atom interaction
in the resonant cavity have been discussed so far
by making the analytic calculations
for the simple case with the constant atom-photon coupling $ \Omega_N $.
They will indeed be confirmed in Sec. \ref{sec:numerical}
by performing detailed numerical calculations
for the realistic situation with the continuous atomic beam
passing through the spatially varying electric field.

\section{Detection sensitivity with continuous atomic beam}
\label{sec:sensitivity}

In order to make precise estimates for the sensitivity
of the Rydberg atom cavity detector, we have to take into accout
(i) the motion and (ii) the almost uniform distribution
of the atoms in the incident beam
as well as (iii) the spatial variation of the electric field in the cavity.
We here elaborate the calculations presented in the preceeding sections
by treating these points appropriately.

The electric field felt by the atoms
varies with time through the atomic motion.
Accordingly, the effect of atomic motion can be incorporated
by introducing the relevant time-dependence for the atom-photon coupling,
which is determined by the profile of the electric field in the cavity,
as given in Eq. (\ref{eqn:Omt}).
On the other hand, in order to treat the spatial distribution of the atoms,
we divide the atoms in the cavity into $ K $ bunches
with a fixed beam intensity
\begin{equation}
I_{\rm Ryd} = N/t_{\rm tr} .
\end{equation}
Here, $ N $ is the total number of Rydberg atoms in the cavity.
Then, the coutinuous atomic beam will actually be realized for $ K \gg 1 $.

The collective atomic mode of each bunch is denoted by $ b_i $
($ i = 1 , 2 , \ldots , K $), and the effective Hamiltonian
is given by a $ (K+2) \times (K+2) $ matrix
\begin{equation}
{\cal H} (t) = \left( \begin{tabular}{c|c}
$ ( \omega_b - \frac{i}{2} \gamma_b ) {\bf 1} $ &
   $ \begin{array}{lc} \Omega_1 (t) & 0 \\
   \vdots & \vdots \\ \Omega_K (t) & 0 \end{array} $ \\
\hline
$ \begin{array}{ccc} \Omega_1 & \ldots & \Omega_K (t) \\
                            0 & \ldots & 0 \end{array} $ &
$ \begin{array}{cc} \omega_c - \frac{i}{2} \gamma_c & \kappa \\
\kappa & \omega_a - \frac{i}{2} \gamma_a
\end{array} $
\end{tabular} \right) .
\label{eqn:HmatrixK}
\end{equation}
Now suppose that the $ i $-th atomic bunch locates around
\begin{equation}
x_i = (i-1) \delta x
\end{equation}
for a time interval
\begin{equation}
M \delta t \leq t < (M+1) \delta t ,
\end{equation}
where $ M = 0 , 1 , 2 , \ldots $, and
\begin{equation}
\delta x = L/K , \
\delta t = t_{\rm tr}/K .
\end{equation}
Then, the collective atom-photon coupling for the $ i $-th bunch
containing $ N/K $ atoms is given
with the electric field profile by
\begin{equation}
\Omega_i (t) = ( \Omega_N / {\sqrt K} ) \ f( x_i + v {\bar t} ) ,
\end{equation}
where
\begin{equation}
{\bar t} \equiv t - M \delta t \ \
( 0 \leq {\bar t} < \delta t ) .
\end{equation}

After each time interval of $ \delta t $,
the $ K $-th atomic bunch leaves the cavity,
and a new one comes in.
Then, the collective atomic modes of the respective bunches
should be replaced as
\begin{equation}
b_i \ \rightarrow \ b_{i+1} ,
\end{equation}
which is represented by a $ ( K + 2 ) \times ( K + 2 ) $ matrix
\begin{equation}
{\cal P}_- = \left( \begin{array}{cccccccc}
0 & 0 & 0 & \cdots & 0 & 0 & 0 & 0 \\
1 & 0 & 0 & \cdots & 0 & 0 & 0 & 0 \\
0 & 1 & 0 & \cdots & 0 & 0 & 0 & 0 \\
\cdots & \cdots & \cdots & \cdots & \cdots & \cdots & \cdots & \cdots \\
0 & 0 & 0 & \cdots & 0 & 0 & 0 & 0 \\
0 & 0 & 0 & \cdots & 1 & 0 & 0 & 0 \\
0 & 0 & 0 & \cdots & 0 & 0 & 1 & 0 \\
0 & 0 & 0 & \cdots & 0 & 0 & 0 & 1
\end{array} \right) .
\end{equation}
In accordance with this replacement of the collective atomic modes
at each period of $ \delta t $,
the time evolution of the particle number matrix $ {\cal N}_{ij} (t) $
should be determined by dividing it into the corresponding parts
in time as
\begin{equation}
{\bar {\cal N}} ({\bar t}, M) \equiv {\cal N} (t) \ \
( M \delta t \leq t < ( M + 1 ) \delta t ) .
\label{eqn:Nbar}
\end{equation}
These parts are connected together at $ t = M \delta t $
with a suitable matching condition
\begin{equation}
{\bar {\cal N}}(0,M+1)
= {\cal P}_- {\bar {\cal N}}( \delta t , M) {\cal P}_-^{\rm T} .
\label{eqn:Nmatch}
\end{equation}
This condition is written down explicitly as
\begin{equation}
\begin{array}{ll}
{\bar {\cal N}}_{ij} (0,M+1)
= {\bar {\cal N}}_{i - 1 \ j -1} ( \delta t , M )
   & [ 2 \leq i , j \leq K ] , \\
{\bar {\cal N}}_{1 j} (0,M+1) = 0
   & [ 1 \leq j \leq K + 2 ] , \\
{\bar {\cal N}}_{i 1} (0,M+1) = 0
   & [ 1 \leq i \leq K + 2 ] , \\
{\bar {\cal N}}_{ij} (0,M+1) = {\bar {\cal N}}_{i - 1 \ j} ( \delta t , M )
   & [ 2 \leq i \leq K , j = a , c ] , \\
{\bar {\cal N}}_{ij} (0,M+1) = {\bar {\cal N}}_{i \ j - 1} ( \delta t , M )
   & [ i = a , c , 2 \leq j \leq K ] , \\
{\bar {\cal N}}_{ij} (0,M+1) = {\bar {\cal N}}_{ij} ( \delta t , M )
   & [ i , j = a , c ] .
\end{array}
\end{equation}
Here, the second and third lines imply
that the incoming atomic mode ($ b_1 $) does not have
any correlation initially with the other quantum modes
($ b_2 , \ldots , b_K , c , a $) already interacting in the cavity.

In this way, by solving the master equation (\ref{eqn:Nmaster})
for many time intervals, we obtain the steady solution
\begin{equation}
{\bar {\cal N}} ({\bar t}) \approx {\bar {\cal N}} ({\bar t}, M \gg 1) .
\end{equation}
It in fact appears independently of the choice of the initial value
$ {\cal N} (0) $, which is due to the dissipation
of the quantum modes for the axions, photons and atoms.
The number of excited atoms contained in each atomic bunch
is then given by
\begin{equation}
n_{b_i} ({\bar t}) = {\bar {\cal N}}_{ii} ({\bar t})
= n^a_{b_i} ({\bar t}) + n^c_{b_i} ({\bar t}) .
\end{equation}
Here, the contributions of the axions and thermal photons
are proporional to the respective particle numbers as
\begin{equation}
n^a_{b_i} ({\bar t}) = r_{b_i a} ({\bar t}) {\bar n}_a , \
n^c_{b_i} ({\bar t}) = r_{b_i c} ({\bar t}) {\bar n}_c .
\label{eqn:nabncb}
\end{equation}
The distribution of excited atoms in the cavity is determined
with this steady solution as
\begin{equation}
{\bar \rho}_b (x) = \frac{n_{b_i} ({\bar t} = x/v )}{\delta x} \ \
((i-1) \delta x \leq x < i \delta x ) .
\label{eqn:nbx}
\end{equation}

The $ K $-th atomic bunch exits the cavity at $ {\bar t} = \delta t $,
and the excited atoms contained there are detected.
Accordingly, the counting rates for the contributions of
the axions and thermal photons are calculated, respectively, 
for large enough $ K $ and $ M $ by
\begin{equation}
R_s = \frac{n^a_{b_K} ( \delta t )}{\delta t} ,
\label{eqn:Rsdet}
\end{equation}
\begin{equation}
R_n = \frac{n^c_{b_K} ( \delta t )}{\delta t} .
\label{eqn:Rndet}
\end{equation}
By using these counting rates, the measurement time required
to search for the axion signal at the confidence level of $ m \sigma $
is estimated as
\begin{equation}
\Delta t = \frac{m^2 ( 1 + R_n / R_s )}{R_s} .
\end{equation}
In the search for the axions with unknown mass,
the cavity frequency ($ \omega_b = \omega_c $ for definiteness)
should be changed with appropriate scanning step $ \Delta \omega_c $.
The total scanning time over a 10\% frequency range
is then given by
\begin{equation}
t_{\rm tot} = \frac{0.1 \omega_c}{\Delta \omega_c} \Delta t .
\end{equation}
The frequency step $ \Delta \omega_c $ is determined
by considering the resonant condition for the absorption
of the axion-converted photons by the Rydberg atoms.
Specifically, as seen in Eqs. (\ref{eqn:OmN1}), (\ref{eqn:sgba1})
and (\ref{eqn:Doma1}), the signal rate is enhanced significantly
for the axion detuning $ | \Delta \omega_a | \lesssim \gamma_a $
with $ \Omega_N / \gamma \sim ( \gamma_a / \gamma )^{1/2} $.
Hence, the scanning frequency step should be taken as
\begin{equation}
\Delta \omega_c \sim \gamma_a .
\label{eqn:Domc}
\end{equation}
Then, if the axion really exists in a mass range
searched with this frequency step,
the resonant condition for the axion signal can be satisfied
at a certain scanning step.

\section{Dependence on the relevant parameters}
\label{sec:dependence}

We here consider how the counting rates of the axion signal
and thermal photon noise depend on the relevant experimental parameters,
before presenting detailed numerical calculations in the next section.
Although the notation for the case of $ K = 1 $
with constant atom-photon coupling $ \Omega_N $ is used for simplicity,
the essential features argued below are indeed valid
for the realistic case with continuous atomic beam.

The energy scales involved in the effective Hamiltonian are as follows:
\[
\begin{array}{lcl}
{\bf detunings} & : & \Delta \omega_a \equiv \omega_a - \omega_c , \
\Delta \omega_b \equiv \omega_b - \omega_c , \\
{\bf couplings} & : & \kappa , \ \Omega_N , \\
{\bf dampings} & : & \gamma_a , \ \gamma_b , \ \gamma_c \equiv \gamma .
\end{array}
\]
The atomic transit time is also important
to determine the time evolution of the system,
which is controlled by the atomic velocity $ v $
with a fixed cavity length $ L $ as
\begin{equation}
v = L / t_{\rm tr} .
\end{equation}
The effects of the dissipation of axions and atoms
are rather small compared to that of the photons
in the actual situation with $ \gamma_a , \gamma_b \ll \gamma $.
The atomic transit time is in fact much longer
than the other characteristic time scales
except for $ \kappa^{-1} $, e.g., $ t_{\rm tr} \simeq 400 \tau_\gamma $
for $ v = 350 {\rm m} {\rm s}^{-1} $
with $ L = 0.2 {\rm m} $, $ Q = 2 \times 10^4 $
and $ m_a = 10^{-5} {\rm eV} $.
In this situation, the quantum averaged particle numbers
approach to the asymptotic values with the factors
$ r_{ij} ( t_{\rm tr} ) \approx r_{ij} ( \infty ) $,
as given in Eq. (\ref{eqn:rij-inf}).
Then, the counting rates of the signal and noise, which are calculated by
$ R_s = r_{ba} ( t_{\rm tr} ) {\bar n}_a / t_{\rm tr} $
and $ R_n = r_{bc} ( t_{\rm tr} ) {\bar n}_c / t_{\rm tr} $,
are found to be roughly proportional to the atomic velocity $ v $.

In the weak atom-photon coupling region of $ \Omega_N / \gamma \ll 0.1 $,
the counting rates $ R_s $ and $ R_n $ are roughly proportional
to $ \Omega_N^2 \propto N \propto I_{\rm Ryd} $,
as seen in Eq. (\ref{eqn:sgba0}) for the axion signal.
On the other hand, when the atom-photon coupling becomes significant
with sufficient atomic beam intensity $ I_{\rm Ryd} $,
the noise rate saturates to certain value
$ R_n \sim {\bar n}_c / t_{\rm tr} $,
and the signal rate $ R_s $ is maximized,
as seen in Eqs. (\ref{eqn:OmN1}) and  (\ref{eqn:sgba1}),
for the atom-photon coupling
\begin{equation}
\Omega_N / \gamma \sim ( \gamma_a / \gamma )^{1/2}
\sim 0.1 \left( \frac{\beta_a}{10^{-3}} \right)
\left( \frac{Q}{10^4} \right)^{1/2} .
\label{eqn:OmNopt}
\end{equation}
It is possible to attain this optimal coupling
in the actual detection apparatus
by preparing the appropriate number of Rydberg atoms
with a suitable laser system.
The beam intensity of Rydberg atoms $ I_{\rm Ryd} = N / t_{\rm tr} $
providing this optimal value (\ref{eqn:OmNopt})
for the collective atom-photon coupling $ \Omega_N = \Omega {\sqrt N} $
is chosen depending on the relevant parameters as
\begin{equation}
I_{\rm Ryd} \sim v \beta_a^2 m_a^2 Q^{-1} \Omega^{-2} .
\label{eqn:IRydopt}
\end{equation}
It is typically estimated as
$ I_{\rm Ryd} \simeq 7 \times 10^5 {\rm s}^{-1} $
with $ N \simeq 400 $ atoms for $ m_a = 10^{-5} {\rm eV} $,
$ \beta_a = 10^{-3} $, $ Q = 2 \times 10^4 $,
$ \Omega = 5 \times 10^3 {\rm s}^{-1} $,
$ v = 350 {\rm m} {\rm s}^{-1} $ and $ L = 0.2 {\rm m} $
($ t_{\rm tr} \simeq 400 \tau_\gamma $).
On the other hand, if the atom-photon coupling becomes too strong,
the signal rate decreases due to the appearance
of Rabi splitting, as seen in Fig. \ref{fig:sgiada1}.

We have observed that the atom-photon interaction can not be treated
perturbatively for $ \Omega_N / \gamma \gtrsim 0.1 $
\cite{Haroche85}.
The conversion between atoms and photons
appears to be reversible in this case.
On the other hand, since the axion-photon coupling is extremely small,
the conversion of axions to photons can well be treated perturbatively
as an irreversible process.
The signal rate $ R_s $ is then found to be proportional
to $ ( \kappa / \gamma )^2 {\bar n}_a $ in a very good approximation.
Then, by considering the relations (\ref{eqn:na}), (\ref{eqn:gamma})
and (\ref{eqn:kappa}) with $ \omega_c \simeq m_a / \hbar $,
the dependence of the signal rate on the relevant parameters
is specified as
\begin{equation}
R_s \sim v c_{a \gamma \gamma}^2 B_{\rm eff}^2 Q^2 V_1 ( \rho_a / m_a )
{\bar \sigma}_{ba} .
\label{eqn:Rs-1}
\end{equation}
Here, the form factor
$ {\bar \sigma}_{ba} = \sigma_{ba} ( \pm \Delta \omega_c / 2 ) $
for $ \Delta \omega_a = \pm \Delta \omega_c / 2 $
($ \Delta \omega_b = 0 $) is also indicated.
This form factor actually depends on the choices
of the atom-photon coupling $ \Omega_N $
and the scanning frequency step $ \Delta \omega_c $,
as seen in Eqs. (\ref{eqn:OmN1}), (\ref{eqn:sgba1}) and (\ref{eqn:Doma1}).
It is noticed in Eq. (\ref{eqn:Rs-1}) that the signal rate is proportional
to the number of axions contained in the conversion cavity
$ ( \rho_a / m_a ) V_1 $.
The dependence of the signal rate on the axion mass $ m_a $
is further specified by considering the relation for
the conversion cavity volume
\begin{equation}
V_1 \sim m_a^{-2} .
\label{eqn:V1-ma}
\end{equation}
This relation is due to the fact that the diameter of the cavity
is proportional to $ m_a^{-1} $ while its length is fixed.

To summarize these arguments,
the optimal situation, as given in
Eqs. (\ref{eqn:OmN1}), (\ref{eqn:sgba1}) and (\ref{eqn:Doma1}),
can be set up experimentally for the dark matter axion search
with Rydberg atom cavity detector.
Then, the counting rates are expected to behave with respect to
the changes of the relevant parameters as
\begin{eqnarray}
R_s & \sim & v c_{a \gamma \gamma}^2 B_{\rm eff}^2 Q
\beta_a^{-2} m_a^{-3} \rho_a ,
\label{eqn:Rsdep} \\
R_n & \sim & v {\bar n}_c .
\label{eqn:Rndep}
\end{eqnarray}
Here, the noise rate is proportional to the thermal photon number
$ {\bar n}_c [ m_a / T_c ] $
which is determined as a function of the ratio $ m_a / T_c $.
The resonant value $ {\bar \sigma}_{ba} \sim ( \gamma_a / \gamma )^{-1}
\sim \beta_a^{-2} Q^{-1} $ is obtained
for the axion-photon-atom interaction
with the small enough axion detuning
$ | \Delta \omega_a | \leq \Delta \omega_c / 2 \sim \gamma_a $.
(This is valid as long as the quantum uncertainty of energy
$ \sim \hbar / t_{\rm tr} $ due to the finite atomic transit time
is smaller than the axion width $ \hbar \gamma_a $,
as ensured in the actual detection system.)
It should also be remarked that
the small detuning $ \Delta \omega_b $
of the atomic transition frequency only shifts slightly
the location of the peak of the axion signal
from $ \Delta \omega_a ({\rm peak}) = 0 $
to $ \Delta \omega_a ({\rm peak}) \simeq \Delta \omega_b $.
By taking these signal and noise rates in the optimal case,
the measurement time $ \Delta t $ at each scanning step
is estimated with respect to the relevant parameters as
\begin{equation}
\Delta t \sim \left\{ \begin{array}{ll}
v^{-1} c_{a \gamma \gamma}^{-4} B_{\rm eff}^{-4}
Q^{-2} \beta_a^4 m_a^6 \rho_a^{-2} {\bar n}_c
& ( R_s / R_n < 1 ) \\
c_{a \gamma \gamma}^{-2} B_{\rm eff}^{-2}
Q^{-2} \beta_a^2 m_a^3 \rho_a^{-1}
& ( R_s / R_n \gtrsim 1 ) \end{array} \right. .
\label{eqn:Dtopt}
\end{equation}
The total scanning time is then estimated
with the appropriate scanning step $ \Delta \omega_c \sim \gamma_a $ as
\begin{equation}
t_{\rm tot} \sim \Delta t / \beta_a^2 .
\label{eqn:ttopt}
\end{equation}
Here the negative power of $ \beta_a $ appearing in the right-hand side
is indeed compensated by the positive power of $ \beta_a $
contained in Eq. (\ref{eqn:Dtopt}) for $ \Delta t $.
The sensitivity seems to be improved
apparently for the high atomic velocity
in Eqs. (\ref{eqn:Dtopt}) and (\ref{eqn:ttopt}).
It should, however, be remarked that the condition
$ \hbar / t_{\rm tr} < \hbar \gamma_a $ for the resonant axion-photon-atom
interaction is no longer ensured if the atomic velocity beccomes too high.
The high atomic beam intensity is even required in such a case
in accordance with Eq. (\ref{eqn:IRydopt}).
Hence, we will in fact see in the next section
that the preferable atomic velocity is
$ v \sim 100 {\rm m} {\rm s}^{-1} - 1000 {\rm m} {\rm s}^{-1} $
for the actual experimental apparatus.

Some essential features have been discussed so far
concerning the detection sensitivity of the Rydberg atom cavity detector.
In the next section, they will indeed be confirmed
by detailed numerical calculations for the realistic situation
with continuous atomic beam.

\section{Numerical analysis}
\label{sec:numerical}

\begin{figure}
\begin{center}
\includegraphics[height=14cm]{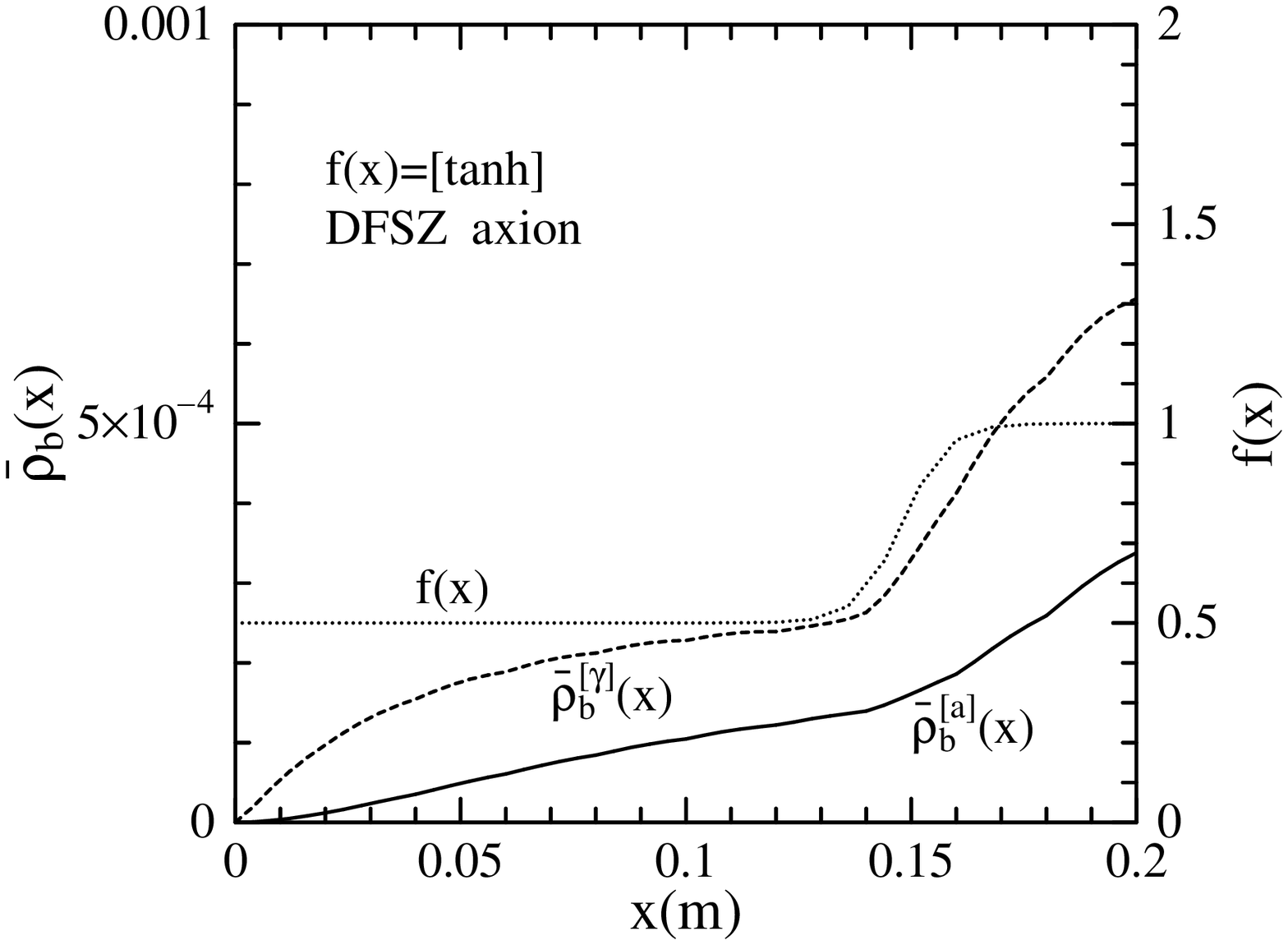}
\caption{The distribution of excited atoms in the cavity
for the tanh-type electric field profile $ f(x) $.
The relevant parameters are taken as
$ m_a = 10^{-5} {\rm eV} $ ($ \omega_c = \omega_a $),
$ T_c = 12 {\rm mK} $, $ Q = 2 \times 10^4 $,
$ \Omega_N / \gamma = 0.1 $ and $ v = 350 {\rm m} {\rm s}^{-1} $.
The contribution of axions $ {\bar \rho}^{[a]}_b (x) $
and that of thermal photons $ {\bar \rho}^{[ \gamma ]}_b (x) $
are shown together.
\label{fig:nbx-t1}}
\end{center}
\end{figure}

\begin{figure}
\begin{center}
\includegraphics[height=14cm]{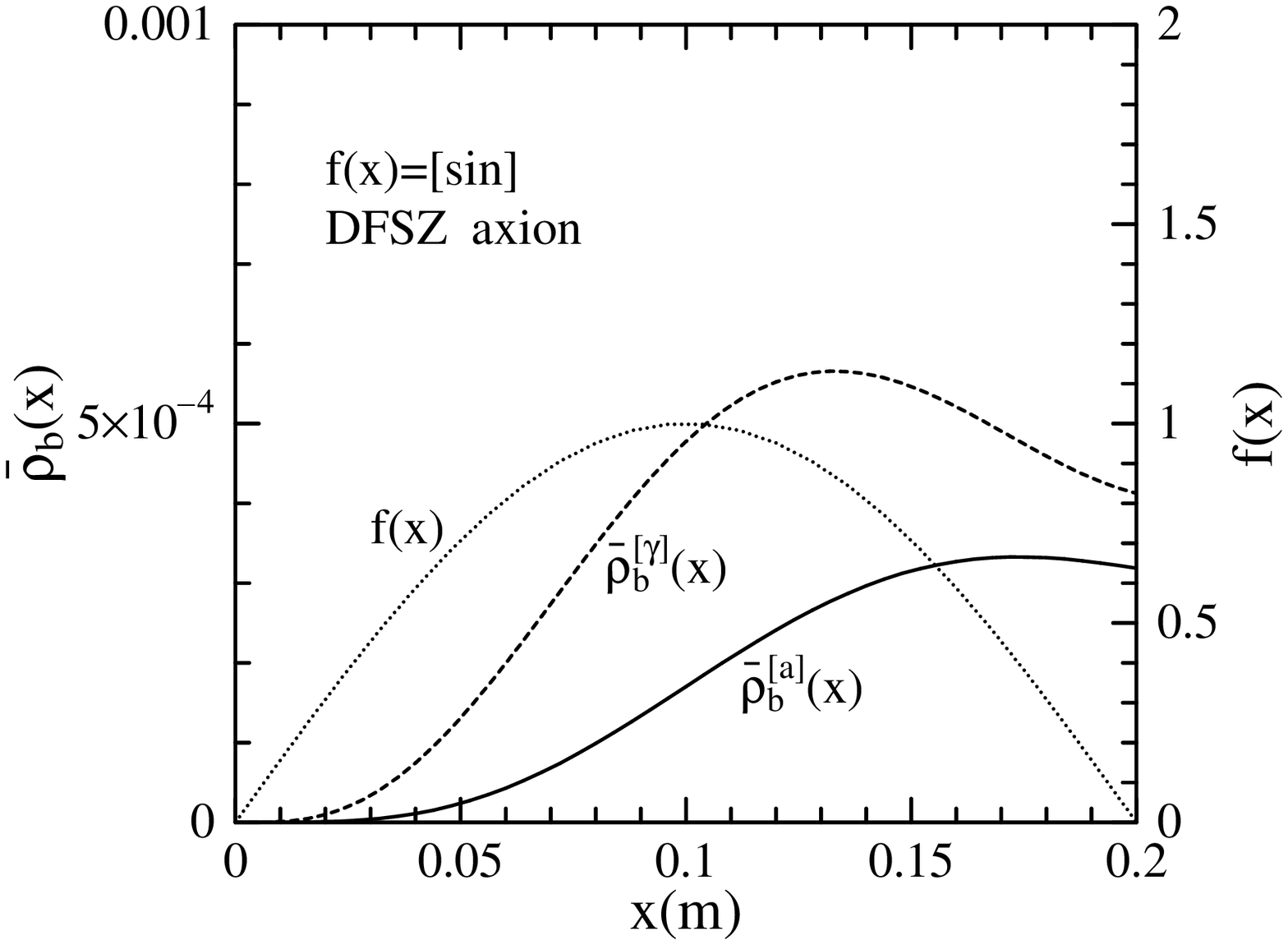}
\caption{The distribution of excited atoms in the cavity
for the sine-type electric field profile $ f(x) $.
The relevant parameters are taken as
$ m_a = 10^{-5} {\rm eV} $ ($ \omega_c = \omega_a $),
$ T_c = 12 {\rm mK} $, $ Q = 2 \times 10^4 $,
$ \Omega_N / \gamma = 0.1 $ and $ v = 350 {\rm m} {\rm s}^{-1} $.
The contribution of axions $ {\bar \rho}^{[a]}_b (x) $
and that of thermal photons $ {\bar \rho}^{[ \gamma ]}_b (x) $
are shown together.
\label{fig:nbx-s1}}
\end{center}
\end{figure}

Numerical calculations have been performed to determine precisely
the quantum evolution of the axion-photon-atom system
in the resonant cavity, where some practical values are taken
for the experimental parameters such as
\[
\begin{array}{l}
m_a = 3 \times 10^{-6}{\rm eV} - 3 \times 10^{-5}{\rm eV} , \\
\beta_a = 3 \times 10^{-4} - 3 \times 10^{-3} , \\
\rho_a = \rho_{\rm halo} = 0.3 {\rm GeV} {\rm cm}^{-3} , \\
T_c = 10{\rm mK} - 15{\rm mK} , \\
Q = 1 \times 10^4 - 7 \times 10^4 , \ B_{\rm eff} = 4 {\rm T} , \\
V_1 = 5000 {\rm cm}^3 \left( m_a / 10^{-5} {\rm eV} \right)^{-2} , \
L = 0.2 {\rm m} , \\
I_{\rm Ryd} = 10^3 {\rm s}^{-1} - 10^7 {\rm s}^{-1} , \\
v = 350 {\rm m} {\rm s}^{-1} - 10000 {\rm m} {\rm s}^{-1} , \\
\Omega = 5 \times 10^3 {\rm s}^{-1} , \ \tau_b = 10^{-3} {\rm s} .
\end{array}
\]
The steady state is realized by solving the master equation
(\ref{eqn:Nmaster}) for a long enough time interval $ M \delta t $
($ M \gg 1 $).
It in practice appears independently of the choice of initial value
$ {\cal N} (0) $ due to the finite damping rates of the axions, photons
and atoms.
The spatial distribution of the excited atoms in the cavity
is calculated by Eq. (\ref{eqn:nbx}) with this steady solution.
The contributions of the galactic axions $ {\bar \rho}_b^{[a]} (x) $
and the thermal photons $ {\bar \rho}_b^{[ \gamma ]} (x) $
are depicted together in Figs. \ref{fig:nbx-t1} and \ref{fig:nbx-s1}
for the case of DFSZ axion with the tanh-type and sine-type
electric field profiles, respectively,
where the relevant parameters are taken as
$ m_a = 10^{-5} {\rm eV} $ ($ \omega_c = \omega_a $), $ T_c = 12 {\rm mK} $,
$ Q = 2 \times 10^4 $, $ \Omega_N / \gamma = 0.1 $
and $ v = 350 {\rm m} {\rm s}^{-1} $.
These results obtained with $ K = 10 $ and $ M = 20 $
in fact appear to be smooth enough indicating
that the steady state in the case with continuous atomic beam
is realized in a good approximation.
It is here remarkable that the form of electric field profile
is reflected in the distribution of the excited atoms.
In the following calculations, we take the tanh-type electric field profile
for definiteness.  The essential results concerning
the interaction between axions and atoms mediated by photons
in the resonant cavity, however, hold even
in the cases with more realistic electric profiles.

We can compute the signal and noise rates with the steady solutions
obtained in this way.
(The following calculations are made for $ K = 5 $ and $ M = 10 $
with the tanh-type electric field profile.
The results are changed less than 10\%
even if larger $ K $ and $ M $ are taken.)
It is readily checked by these calculations
that the signal rate $ R_s $ is indeed proportional to
$ c_{a \gamma \gamma}^2 B_{\rm eff}^2 V_1 ( \rho_a / m_a ) $,
as shown in Eq. (\ref{eqn:Rs-1}).
Hence, the signal rate for the KSVZ axion is larger than
that for the DFSZ axion by a factor
$ c_{a \gamma \gamma}^2 ({\rm KSVZ}) / c_{a \gamma \gamma}^2 ({\rm DFSZ})
\simeq 7 $.

\begin{figure}
\begin{center}
\includegraphics[height=14cm]{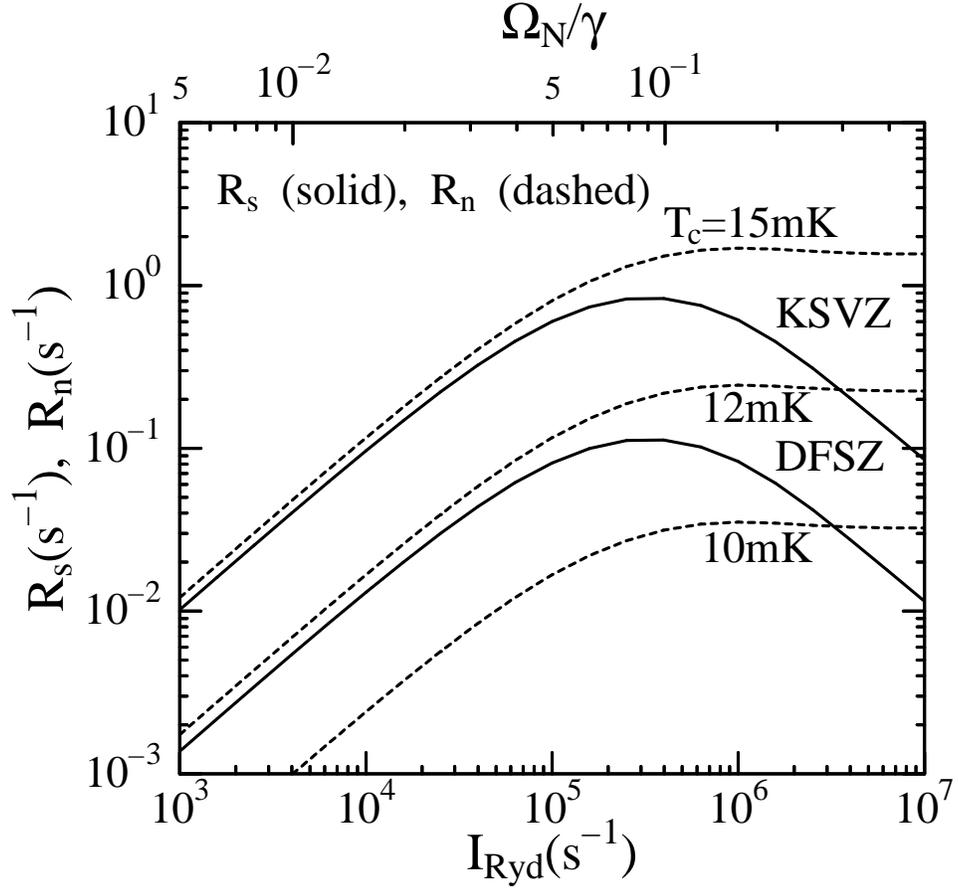}
\caption{The signal and noise rates versus the atomic beam intensity
(atom-photon coupling).
The relevant parameters are taken as
$ m_a = 10^{-5} {\rm eV} $ ($ \omega_c = \omega_a $), $ Q = 2 \times 10^4 $,
$ L = 0.2 {\rm m} $, $ v = 350 {\rm m} {\rm s}^{-1} $
and $ \Omega = 5 \times 10^3 {\rm s}^{-1} $.
\label{fig:rsrnitq1}}
\end{center}
\end{figure}

In Fig. \ref{fig:rsrnitq1}, we plot the signal and noise rates,
$ R_s $ and $ R_n $, together depending on the atomic beam intensity
$ I_{\rm Ryd} $
(and the corresponding atom-photon coupling $ \Omega_N $),
where the relevant parameters are taken as
$ m_a = 10^{-5} {\rm eV} $ ($ \omega_c = \omega_a $),
$ Q = 2 \times 10^4 $, $ L = 0.2 {\rm m} $,
$ v = 350 {\rm m} {\rm s}^{-1} $, $ \Omega = 5 \times 10^3 {\rm s}^{-1} $
and $ T_c = 10 {\rm mK} , \ 12 {\rm mK} , \ 15 {\rm mK} $.
The noise rate is clearly proportional to
the thermal photon number $ {\bar n}_c [ m_a / T_c ] $,
which is $ {\bar n}_c = 8.9 \times 10^{-6} , \ 6.2 \times 10^{-5} , \
4.3 \times 10^{-4} $
at $ T_c = 10 {\rm mK} $, $ 12 {\rm mK} $, $ 15 {\rm mK} $, respectively,
for the axion mass $ m_a = 10^{-5} {\rm eV} $.
In the weak beam intensity region, the signal and noise rates
increase almost proportional to $ I_{\rm Ryd} = N / t_{\rm tr} $.
On the other hand, for sufficiently strong beam intensities
the noise rate is saturated to certain asymptotic value.
In this case, the atoms interact reversibly with the photons
in the cavity so that the equilibrium
with $ n_b [c \rightarrow b] \simeq {\bar n}_c $ is realized.
(In practice, due to the finite atomic lifetime
the total number of excited atoms $ n_b [c \rightarrow b]
= \int_0^L {\bar \rho}_b^{[ \gamma ]} (x) dx $
obtained from the thermal photons is slightly different from
$ {\bar n}_c $.)
As for the signal rate, we clearly observe in Fig. \ref{fig:rsrnitq1}
that it is optimized for certain atomic beam intensity
corresponding to the condition (\ref{eqn:OmNopt})
for the atom-photon coupling.
If the atomic beam intensity is too strong,
the signal rate even decreases due to the Rabi splitting.
It is hence important to adjust the beam intensity
so as to attain the optimal condition for the signal.

It is also checked that the lines representing the noise rate $ R_s $
versus beam intensity $ I_{\rm Ryd} $ shifts horizontally
by changes of the intrinsic atom-photon coupling $ \Omega $.
Although the theoretical estimate
of the electric dipole transition matrix element $ d $ is available
\cite{DZ81}
for calculating $ \Omega $ with Eq. (\ref{eqn:Omega}),
some ambiguities would be present
in such a naive theoretical estimate of $ \Omega $.
Hence, it may rather be necessary to determine $ \Omega $ experimentally
by fitting the thermal noise data in the weak beam intensity region
to the expected lines, as shown in Fig. \ref{fig:rsrnitq1}
for $ \Omega = 5 \times 10^3 {\rm s}^{-1} $,
which are calculated by varying $ \Omega $ around some plausible value
suggested by the theoretical calculation of $ d $
\cite{DZ81}.

Another aspect should be pointed out concerning the counting rates
of the excited atoms.
As given in Eqs. (\ref{eqn:nbx}) and (\ref{eqn:Rndet}),
the noise rate is calculated by the formula
$ R_n = v {\bar \rho}_b^{[ \gamma ]} (L) $
with the excited atom density around the exit of cavity ($ x = L $).
Then, it may be noticed in Fig. \ref{fig:rsrnitq1}
that the saturated value of $ R_n $ in the strong beam intensity region,
e.g., $ 0.23 {\rm s}^{-1} $ for $ T_c = 12 {\rm mK} $,
is in fact significantly larger than the naive estimate
$ {\bar n}_c / t_{\rm tr} $, e.g., $ 0.11 {\rm s}^{-1} $
for $ {\bar n}_c [ 10^{-5}{\rm eV}/12 {\rm mK} ] = 6.2 \times 10^{-5} $,
$ v = 350 {\rm m} {\rm s}^{-1} $ and $ L = 0.2 {\rm m} $.
This enhancement of the counting rates,
which is expected for the signal as well as the noise,
is due to the fact that the densities of excited atoms
$ {\bar \rho}_b^{[a]} (L) $ and $ {\bar \rho}_b^{[ \gamma ]} (L) $
around the exit of cavity become higher than the average values,
as seen in Figs. \ref{fig:nbx-t1} and \ref{fig:nbx-s1}.
The nonuniform density of the excited atoms in the cavity
is indeed brought by the fact that the atoms in the continuous beam
are passing through the spatially varying electric field
with finite transit time.

\begin{figure}
\begin{center}
\includegraphics[height=14cm]{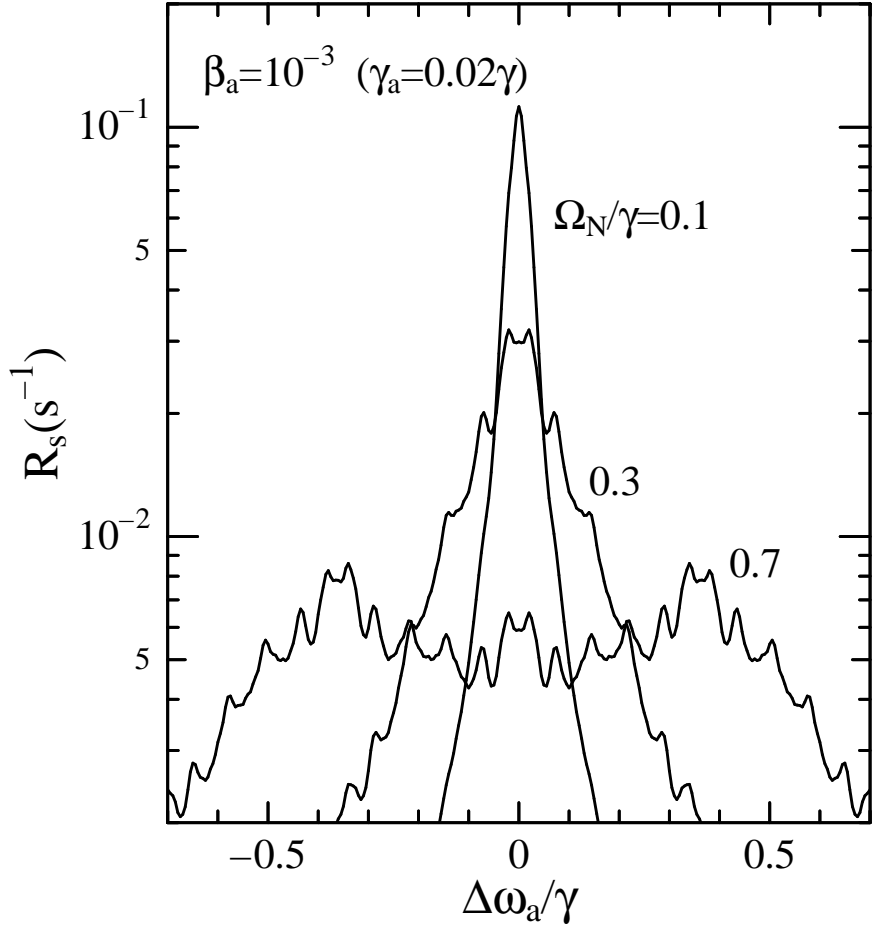}
\caption{The counting rate of signal depending on the axion detuning
for various values of the atom-photon coupling $ \Omega_N $.
The relevant parameters are taken as
$ m_a = 10^{-5} {\rm eV} $, $ \beta_a = 10^{-3} $, $ Q = 2 \times 10^4 $,
$ L = 0.2 {\rm m} $ and $ v = 350 {\rm m} {\rm s}^{-1} $.
\label{fig:rsda-o1}}
\end{center}
\end{figure}

The behavior of the signal rate $ R_s $ with respect to the axion detunig
$ \Delta \omega_a $ is shown in Fig. \ref{fig:rsda-o1}.
Here some typical values are taken for the atom-photon coupling,
$ \Omega_N / \gamma = 0.1 , \ 0.3 , \ 0.7  $,
and the other relevant parameters are chosen as
$ m_a = 10^{-5} {\rm eV} $, $ \beta_a = 10^{-3} $,
$ Q = 2 \times 10^4 $, $ L = 0.2 {\rm m} $
and $ v = 350 {\rm m} {\rm s}^{-1} $.
This behavior almost agrees with that of the form factor
in Fig. \ref{fig:sgiada1} which is obtained for the simple case.
(Note that the log-scale is taken for $ R_s $ in Fig. \ref{fig:rsda-o1}.)
It is therefore confirmed by these numerical calculations
that the resonant axion-photon-atom interaction
takes place if the axion detuning is small enough as
$ | \Delta \omega_a | \lesssim \gamma_a $
with the optimal atom-photon coupling
$ \Omega_N / \gamma \sim ( \gamma_a / \gamma )^{1/2} $
(as long as $ \hbar / t_{\rm tr} < \hbar \gamma_a $).
It is also checked that if there is a small detuning $ \Delta \omega_b $
of the atomic transition frequency, this resonant condition
is slightly modified as $ | \Delta \omega_a - \Delta \omega_b |
= | \omega_a - \omega_b | \lesssim \gamma_a $.

As clearly observed in  Fig. \ref{fig:rsda-o1},
a salient feature which is found by these calculations
for the case with continous atomic beam is the appearance
of a ripple structure in $ R_s $ versus $ \Delta \omega_a $
with the relatively strong atom-photon coupling,
$ \Omega_N / \gamma > 0.1 $ for the present choice of parameters.
It is realized that this structure is brought
from the narrow axion width $ \gamma_a \sim \beta_a^2 m_a / \hbar $.
We have checked that this fine structure indeed disappears
if the axion spectrum spreads much wider with higher mean velocity
such as $ \beta_a = 3 \times 10^{-3} $.

\begin{figure}
\begin{center}
\includegraphics[height=14cm]{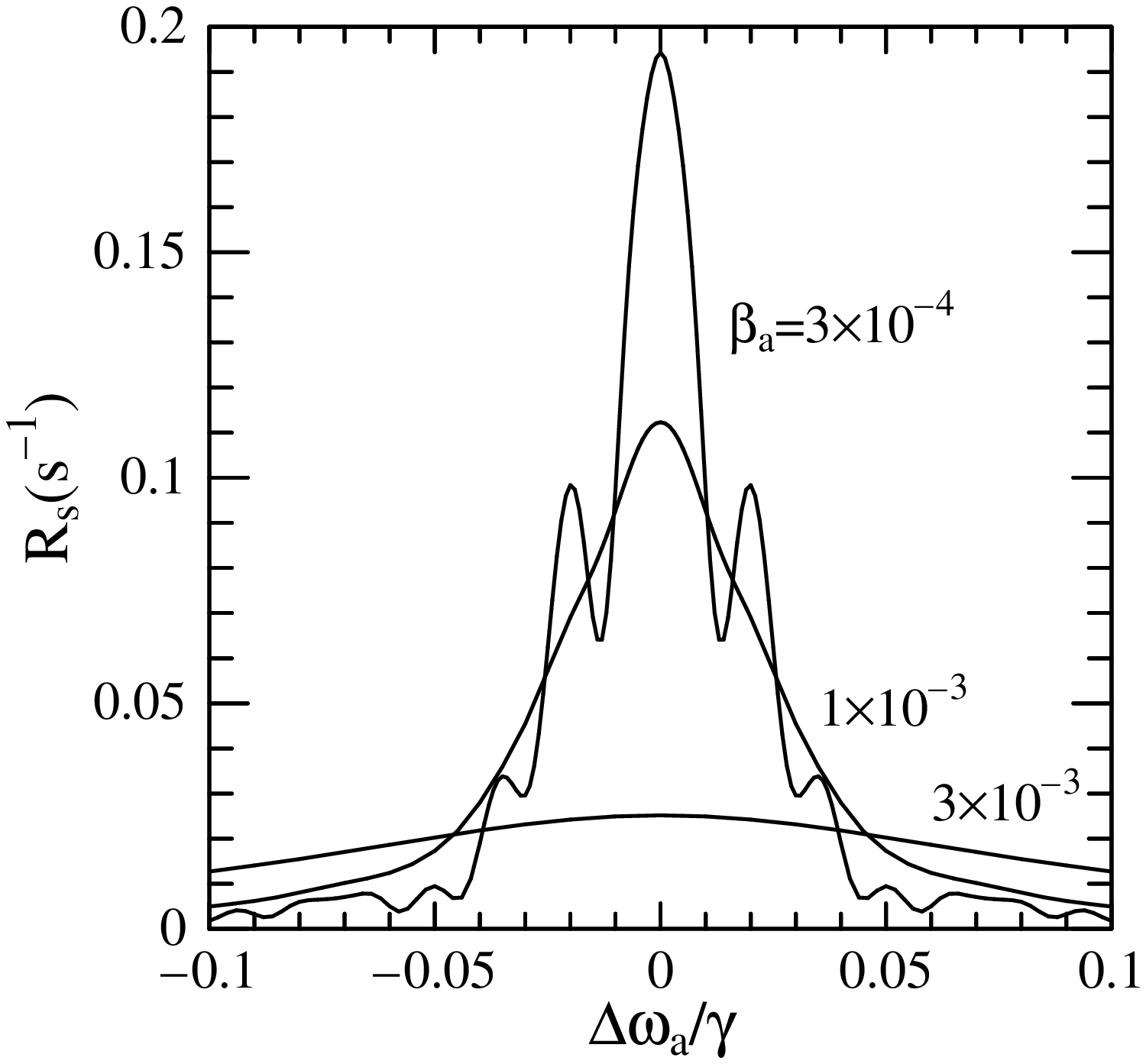}
\caption{The counting rate of signal versus the axion detuning
for various values of the axion velocity $ \beta_a $.
The relevant parameters are taken as
$ m_a = 10^{-5} {\rm eV} $, $ Q = 2 \times 10^4 $,
$ \Omega_N / \gamma = 0.1 $,
$ L = 0.2 {\rm m} $ and $ v = 350 {\rm m} {\rm s}^{-1} $.
\label{fig:rsda01bt}}
\end{center}
\end{figure}

The signal rate $ R_s $ versus axion detuning $ \Delta \omega_a $
is calculated for some typical axion velocities,
$ \beta_a = 3 \times 10^{-4} , 1 \times 10^{-3} , 3 \times 10^{-3} $.
The results are shown in Fig. \ref{fig:rsda01bt},
where $ m_a = 10^{-5} {\rm eV} $, $ Q = 2 \times 10^4 $,
$ \Omega_N / \gamma = 0.1 $, $ L = 0.2 {\rm m} $
and $ v = 350 {\rm m} {\rm s}^{-1} $ are taken.
It is here clearly observed that this signal form factor
exhibits the structure determined by the energy spread
of axions $ \hbar \gamma_a \sim \beta_a^2 m_a $.
The galactic axion spectrum can actually have some narrow peaks,
as pointed out in \cite{Sikivie92}.
Then, by taking a small enough scanning frequency step $ \Delta \omega_c $,
such peaks may be observed in the present detection scheme
as well as the heterodyne method
\cite{Hagmann98}.
The energy resolution is now limited by the qauntum uncertainty
$ \sim \hbar / t_{\rm tr} $, which could be improved
by utilizing the lower velocity atomic beam.

\begin{figure}
\begin{center}
\includegraphics[height=14cm]{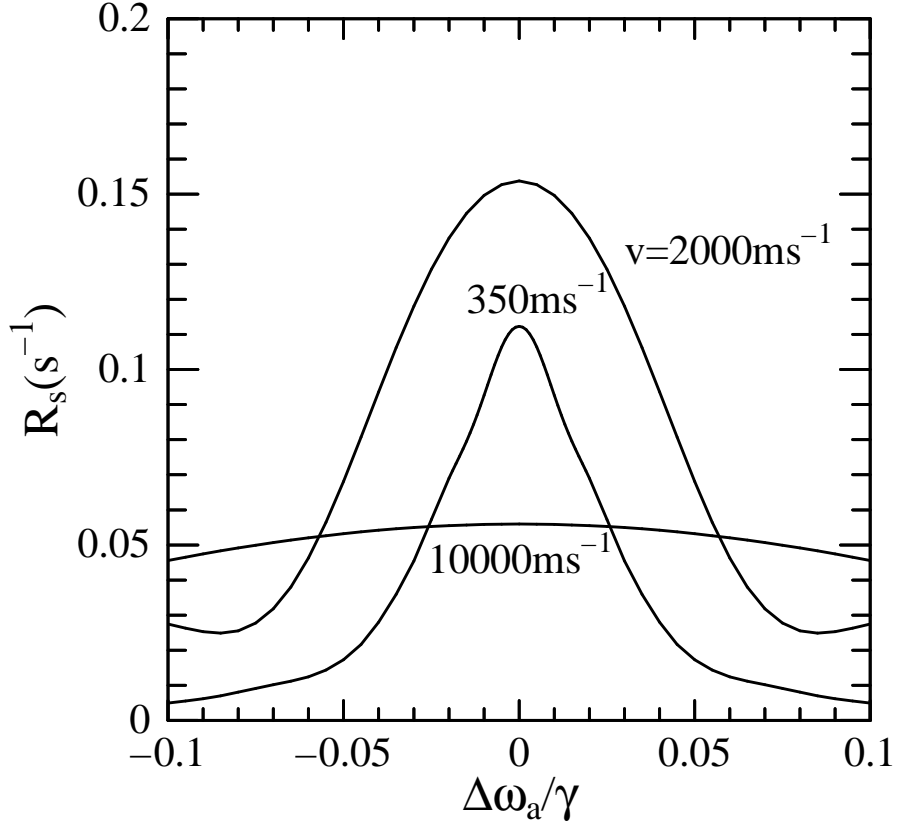}
\caption{The counting rate of signal versus the axion detuning
for various values of the atomic velocity $ v $.
The relevant parameters are taken as
$ m_a = 10^{-5} {\rm eV} $, $ \beta_a = 10^{-3} $,
$ Q = 2 \times 10^4 $, $ \Omega_N / \gamma = 0.1 $ and $ L = 0.2 {\rm m} $.
\label{fig:rsdav-dt}}
\end{center}
\end{figure}

The dependence of the axion signal $ R_s $ on the atomic velocity $ v $
is also shown in Fig. \ref{fig:rsdav-dt}
for $ m_a = 10^{-5} {\rm eV} $, $ \beta_a = 10^{-3} $,
$ \Omega_N / \gamma = 0.1 $, $ L = 0.2 {\rm m} $
and $ v = 350 {\rm m} {\rm s}^{-1} , 2000 {\rm m} {\rm s}^{-1} ,
10000 {\rm m} {\rm s}^{-1} $.
Here, we note that for high atomic velocities such as
$ v = 10000 {\rm m} {\rm s}^{-1} $ this axion signal form factor
has a width much larger than that of axions.
It is in fact determined by the energy uncertainty
$ \sim \hbar / t_{\rm tr} $ (e.g., $ \sim 0.1 \hbar \gamma $
for $ v = 10000 {\rm m} {\rm s}^{-1} $) due to the atomic motion.

\begin{figure}
\begin{center}
\includegraphics[height=14cm]{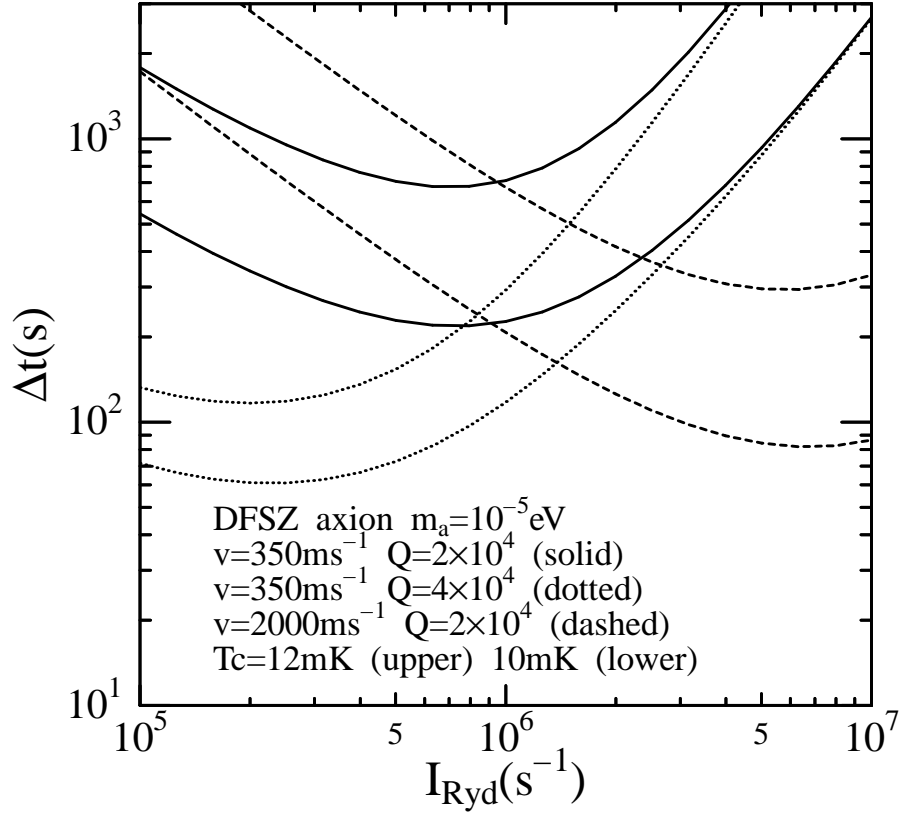}
\caption{The measurment time for $ 3 \sigma $ at each frequency step
depending on the atomic beam intensity,
which is estimated in the case of DFSZ axion.
Here, some typical values are taken for $ T_c $, $ Q $ and $ v $,
and the other relevant parameters are chosen as
$ m_a = 10^{-5} {\rm eV} $, $ \beta_a = 10^{-3} $,
$ \Omega_N / \gamma = 0.1 $,
$ L = 0.2 {\rm m} $ and $ v = 350 {\rm m} {\rm s}^{-1} $.
\label{fig:dtiqv-1}}
\end{center}
\end{figure}

\begin{figure}
\begin{center}
\includegraphics[height=14cm]{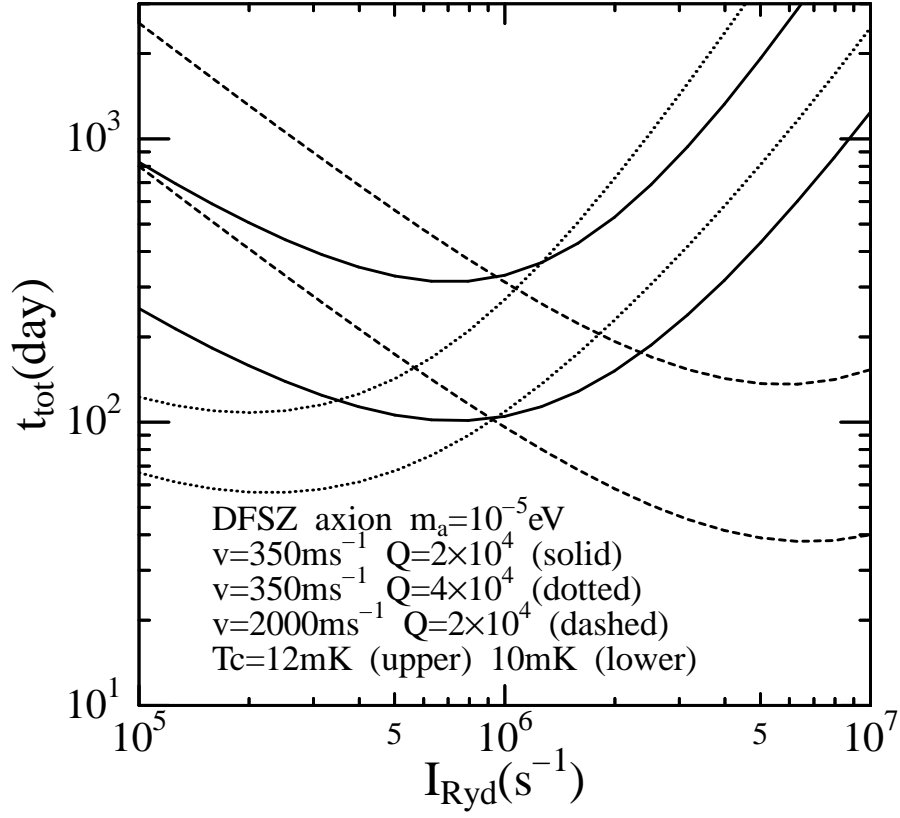}
\caption{The total scanning time ($ 3 \sigma $ level) for the DFSZ axion
depending on the atomic beam intensity.
Here, some typical values are taken for $ T_c $, $ Q $ and $ v $,
and the other relevant parameters are chosen as
$ m_a = 10^{-5} {\rm eV} $, $ \beta_a = 10^{-3} $,
$ \Omega_N / \gamma = 0.1 $,
$ L = 0.2 {\rm m} $ and $ v = 350 {\rm m} {\rm s}^{-1} $.
\label{fig:ttiqv-1}}
\end{center}
\end{figure}

The measurement time $ \Delta t $ required for $ 3 \sigma $ level
at each scanning step is shown in Fig. \ref{fig:dtiqv-1}
depending on the atomic beam intensity $ I_{\rm Ryd} $.
Here the axion detuning is taken for definiteness
as $ | \Delta \omega_a | \leq \Delta \omega_c / 2 $
with the scanning frequency step
\begin{equation}
\Delta \omega_c = 0.05 \gamma
= 6 {\rm kHz} \left( \frac{m_a}{10^{-5}{\rm eV}} \right)
\left( \frac{2 \times 10^4}{Q} \right) ,
\end{equation}
which is of the order of axion width $ \gamma_a $
for $ \beta_a \sim 10^{-3} $
meeting the resonant condition (\ref{eqn:Doma1}) for the axion signal.
Then, the total scanning time $ t_{\rm tot} $ is estimated, as shown
in Fig. \ref{fig:ttiqv-1}.
These calculations are made for the DFSZ axion
by taking some typical values of $ T_c $, $ Q $ and $ v $,
which are indicated in Figs. \ref{fig:dtiqv-1} and \ref{fig:ttiqv-1}.
The other relevant parameters are chosen as
$ m_a = 10^{-5} {\rm eV} $, $ \beta_a = 10^{-3} $,
$ \Omega_N / \gamma = 0.1 $, $ L = 0.2 {\rm m} $
and $ v = 350 {\rm m} {\rm s}^{-1} $.
Here, we can see that the beam intensity to optimize
the detection sensitivity changes with respect to $ Q $ and $ v $
according to the relation (\ref{eqn:IRydopt}), as discussed before.
The optimal estimates of the measurement time
and the total scanning time also show roughly the dependence
on $ Q $ and $ v $ as indicated
in (\ref{eqn:Dtopt}) and (\ref{eqn:ttopt}).
For the KSVZ axion, the sensitivity is much better
by a factor $ \simeq 7^2 \simeq 50 $ ($ R_s / R_n < 1 $)
or $ \simeq 7 $ ($ R_s / R_n \gtrsim 1 $).

\begin{figure}
\begin{center}
\includegraphics[height=14cm]{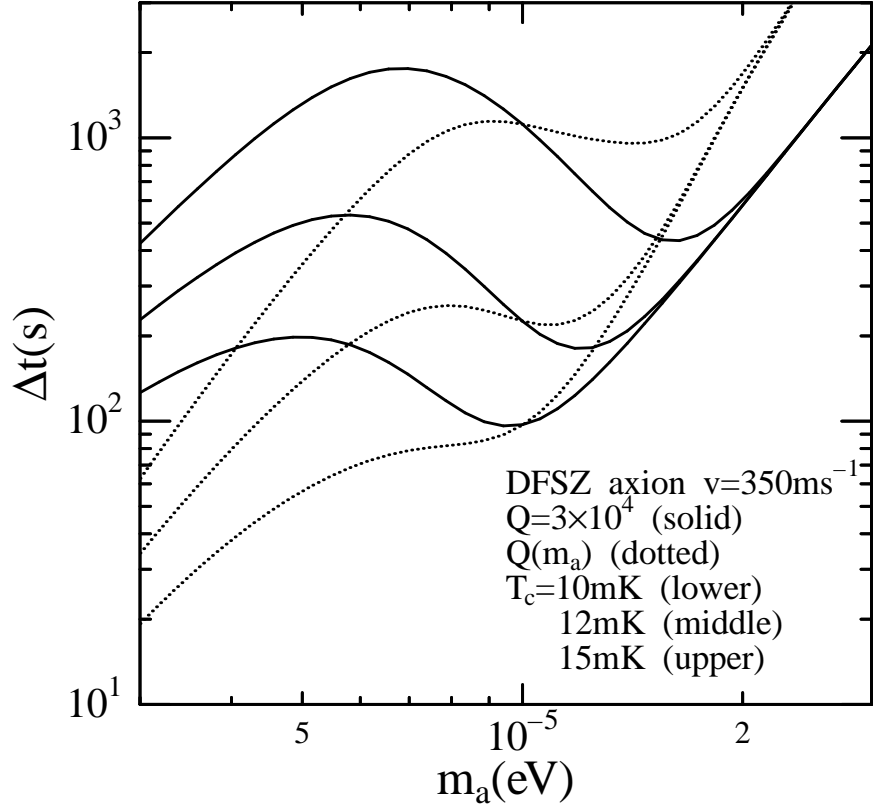}
\caption{The measurment time for $ 3 \sigma $ at each frequency step
depending on the axion mass, which is estimated in the case of DFSZ axion.
The $ Q $ factor is fixed (solid lines)
or varies with the axion mass (dotted lines).
\label{fig:dtmaq-1}}
\end{center}
\end{figure}

\begin{figure}
\begin{center}
\includegraphics[height=14cm]{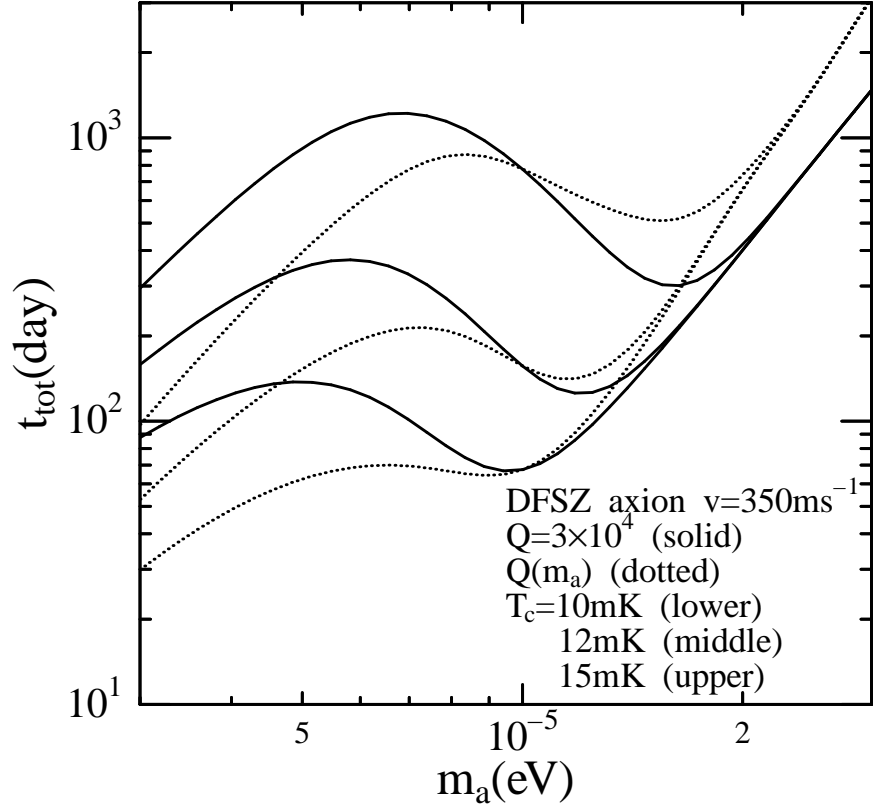}
\caption{The total scanning time ($ 3 \sigma $ level) for the DFSZ axion
depending on the axion mass.
The $ Q $ factor is fixed (solid lines)
or varies with the axion mass (dotted lines).
\label{fig:ttmaq-1}}
\end{center}
\end{figure}

The detection sensitivity for $ 3 \sigma $ is finally estimated
over the axion mass range
$ m_a = 3 \times 10^{-6} {\rm eV} - 3 \times 10^{-5} {\rm eV} $,
which can be searched with the present type of Rydberg atom cavity detector.
In these calculations, the optimal atomic beam intensity
is taken as
$ I_{\rm Ryd} = 4 \times 10^5 {\rm s}^{-1}
( m_a / 10^{-5} {\rm eV} )^2 ( Q / 3 \times 10^4 )^{-1} $
in accordance with the relation (\ref{eqn:IRydopt}),
which is also read from Figs. \ref{fig:dtiqv-1} and \ref{fig:ttiqv-1}.
The volume of the conversion cavity is taken as
$ V_1 = 5000 {\rm cm}^3 ( m_a / 10^{-5} {\rm eV} )^{-2} $
by considering the relation (\ref{eqn:V1-ma}).
The other relevant parameters are chosen as
$ \beta_a = 10^{-3} $, $ L = 0.2 {\rm m} $, $ v = 350 {\rm m}{\rm s}^{-1} $,
$ \Delta \omega_c = 0.05 \gamma $
and $ T_c = 10 {\rm mK} , \ 12 {\rm mK} , \ 15 {\rm mK} $.
The results are shown in Figs. \ref{fig:dtmaq-1} and \ref{fig:ttmaq-1}.
It should here be noted that the $ Q $ factor can actually
increase for lower frequencies.
In order to take into account this property of the $ Q $ factor,
we have assumed for example a relation
\begin{equation}
Q ( m_a ) = 3 \times 10^4 ( 10^{-5} {\rm eV} / m_a )^{2/3} ,
\label{eqn:Qma}
\end{equation}
providing $ Q(m_a) \simeq 1.4 \times 10^4 - 6.7 \times 10^4 $
for $ m_a = 3 \times 10^{-5} {\rm eV} - 3 \times 10^{-6} {\rm eV} $.
The results obtained by taking a fixed value and this relation
for the $ Q $ factor are plotted toghether by solid and dotted lines,
respectively, in Figs. \ref{fig:dtmaq-1} and \ref{fig:ttmaq-1}.

Here we clearly find that if the $ Q $ factor increases
as given in Eq. (\ref{eqn:Qma}), the detection sensitivity can be improved
significantly for the lower axion masses $ m_a \sim 10^{-6} {\rm eV} $.
It is also observed in most cases that the the measurement time $ \Delta t $
and the total scanning time $ t_{\rm tot} $ once become local minimum
for certain axion mass around $ m_a = 10^{-5} {\rm eV} $.
As the axion mass gets smaller from this minimum point,
$ \Delta t $ and $ t_{\rm tot} $ increase
until reaching a local maximum, and then turn to decrease again.
This feature is understood by noting the behaviors
of the signal and noise rates with respect to the axion mass.
In fact, as shown in Eq. (\ref{eqn:Rsdep}), the singal rate increases
approximately as $ R_s \sim m_a^{-3} $ (with fixed $ Q $)
when the axion mass gets smaller.
This increase of the signal rate is eventually overwhelmed
by the more rapid increase of the noise rate
$ R_n \sim {\rm e}^{- m_a / k_{\rm B}T_c} $ for $ m_a / k_{\rm B}T_c \gg 1 $
proportional to the thermal photon number $ {\bar n}_c $.
Then, as the axion mass gets smaller to be comparable
to the cavity temperature $ T_c $,
the increase of the noise rate becomes gradually moderate,
and the singnal rate begins to dominate again.

We can now conclude with these detailed numerical calculations.
If the galactic axion search is made by utilizing
the Rydberg atom cavity detecter,
the DFSZ axion limit can be reached in frequency ranges
of 10\% around the axion mass
$ m_a \sim 10^{-6} {\rm eV} - 10^{-5} {\rm eV} $
for reasonable scanning times.
The optimal condition for the detection sensitivity is attained
by cooling the cavity down to a temperature $ T_c \sim 10 {\rm mK} $
and adjusting the experimental parameters such as
the atomic beam intensity $ I_{\rm Ryd} $ and velocity $ v $
and also the scanning frequency step $ \Delta \omega_c $.

\section{Summary}
\label{sec:sammary}

We have developed quantum theoretical calculations
on the dynamical system consisting
of the cosmic axions, photons and Rydberg atoms which are interacting
in the resonant cavity.
The time evolution is determined for the number of Rydberg atoms
in the upper state which are excited
by absorbing the axion-converted and thermal background photons.
The calculations are made, in particular, by taking into account
the actual experimental situation such as the motion and uniform
distribution of the Rydberg atoms in the incident beam
and also the spatial variation of the electric field in the cavity.
Some essential aspects on the axion-photon-atom interaction
in the resonant cavity are clarified by these detailed calculations.
Then, by using these results the detection sensitivity
of the Rydberg atom cavity detector is estimated properly.
The present quantum analysis clearly shows
that the Rydberg atom cavity detector is quite efficient
for the dark matter axion search.

\acknowledgments

This research was partly supported
by a Grant-in-Aid for Specially Promoted Research
by the Ministry of Education, Science, Sports and Culture,
Japan under the program No. 09102010.

\appendix

\section{Analytic solution}
\label{appendix:analytic}

Basic properties of the axion-photon-atom interaction in the cavity
can be examined by considering the simple case
where the atom-photon coupling does not depend on the time
(i.e., the spatial variation of the electric field in the cavity
is not considered).
In this case, the analytic solution is obtained
for the particle number matrix $ {\cal N}_{ij} (t) $
as follows \cite{OMY96}.

We first consider a more general case with $ K $ atomic bunches
in the incident beam where the atom-photon couplings
$ \Omega_1 $ -- $ \Omega_K $, which may be defferent each other,
are independent of the time.
The effective Hamiltonian $ {\cal H} $ given
in Eq. (\ref{eqn:HmatrixK})
with constant $ \Omega_1 $ -- $ \Omega_K $ is diagonalized
by a nonsingular linear transformation $ {\cal T} $
(not unitary for nonhermitian $ {\cal H} $):
\begin{equation}
{\bar {\cal H}} = {\cal T}^{-1} {\cal H} {\cal T}
= {\rm diag.} ( \lambda_1 , \lambda_2 , \ldots, \lambda_{K+2} ) .
\end{equation}
Accordingly, the relevant matrices are introduced by
\begin{eqnarray}
{\bar {\cal N}}^\prime (t)
&=& {\cal T}^{-1 *} {\cal N}^\prime (t) {\cal T}^{-1 {\rm T}} , \\
{\bar {\cal D}}
&=& {\cal T}^{-1 *} {\cal D} {\cal T}^{-1 {\rm T}} ,
\end{eqnarray}
where $ {\cal N}^\prime (t)
= {\cal U}^{-1 *} (t) {\cal N}(t) {\cal U}^{-1 {\rm T}} (t) $,
as given in Eq. (\ref{eqn:Nprmt}),
with $ {\cal U} (t) = \exp [ - i {\cal H} t ] $
for the time-independent $ {\cal H} $.
Then, the master equation (\ref{eqn:Nmaster}) is reduced
to a set of unmixed equations,
\begin{equation}
\frac{d {\bar {\cal N}}^\prime_{ij}}{dt}
= {\rm e}^{-i ( \lambda_i^* - \lambda_j ) t} {\bar {\cal D}}_{ij} .
\end{equation}
These equations are readily solved with an appropriate initial value
$ {\cal N} (0) = {\cal N}^\prime (0) $ as
\begin{equation}
{\bar {\cal N}}^\prime_{ij} (t)
= {\bar {\cal N}}^\prime_{ij} (0) + {\bar {\cal C}}_{ij} (t) ,
\label{eqn:Nbarprmt}
\end{equation}
where
\begin{equation}
{\bar {\cal C}}_{ij} (t)
= i \frac{{\bar {\cal D}}_{ij}}{\lambda_i^* - \lambda_j}
\left[ {\rm e}^{-i ( \lambda_i^* - \lambda_j ) t} - 1 \right] .
\label{eqn:Dltbarana}
\end{equation}
By using this solution for $ {\bar {\cal N}}^\prime (t) $,
the original $ {\cal N} (t) $ is caluclated as
\begin{equation}
{\cal N}(t) = {\cal U}^* (t)
\left[ {\cal N}(0) + {\cal C} (t) \right] {\cal U}^{\rm T} (t) ,
\label{eqn:Ntana}
\end{equation}
where
\begin{eqnarray}
{\cal C} (t) &=& {\cal T}^* {\bar {\cal C}} (t) {\cal T}^{\rm T} , \\
\label{eqn:Dltana}
{\cal U}(t) &=& {\cal T} \exp [ -i {\bar {\cal H}} t ] {\cal T}^{-1} .
\label{eqn:Utana}
\end{eqnarray}
The above procedure with Eqs. (\ref{eqn:Nbarprmt}) -- (\ref{eqn:Utana})
to find the solution may be viewed
as an inhomogeneous linear transformation from $ {\cal N} (0) $
to $ {\cal N}(t) $ depending on the form of $ {\cal H} $:
\begin{equation}
{\cal N}(t) = {\cal {\cal U}}_{\cal H} [ {\cal N}(0) ] .
\label{eqn:UHN}
\end{equation}

The analytic solution obtained in this way can be used
in a good approximation for a sufficiently small time interval
where the variation of the atom-photon coupling is neglected.
(This is equivalent to approximate the electric field profile $ f(x) $
with a relevant step-wise function, if the time interval is taken
to be $ \delta t = \delta x / v $.)
Then, by applying the transformation (\ref{eqn:UHN})
for a time sequence $ t \equiv t_{z+1} > t_z > t_{z-1} > \cdots > t_0 $
with small intervals $ t_k - t_{k-1} $ ($ 1 \leq k \leq z+1 $)
the approximate solution for $ {\cal N} (t) $ may be obtained
as
\begin{equation}
{\cal N} (t_0)
\stackrel{\displaystyle{{\cal U}_{{\cal H}(t_0)}}}{ \longrightarrow}
{\cal N} (t_1)
\stackrel{\displaystyle{{\cal U}_{{\cal H}(t_1)}}}{ \longrightarrow}
\cdots \stackrel{\displaystyle{{\cal U}_{{\cal H}(t_z)}}}{ \longrightarrow}
{\cal N}(t) .
\end{equation}

We now consider the specific case of $ K = 1 $
with $ \Omega (t) = \Omega_N $, where the Hamiltonian is given by
\begin{equation}
{\cal H} = \left( \begin{array}{ccc}
\omega_b - \frac{i}{2} \gamma_b & \Omega_N & 0 \\
\Omega_N & \omega_c - \frac{i}{2} \gamma_c & \kappa \\
0 & \kappa & \omega_a - \frac{i}{2} \gamma_a  \\
\end{array} \right) .
\label{eqn:Hmatrix0}
\end{equation}
Then, the transformation matrix $ {\cal U}(t) $ is calculated
from Eq. (\ref{eqn:Utana}) as
\begin{equation}
{\cal U}_{ij} (t) = \sum_k g^k_{ij} {\rm e}^{-i \lambda_k t} ,
\end{equation}
with
\begin{equation}
g^k_{ij} = {\cal T}_{ik} {\cal T}^{-1}_{kj} ,
\end{equation}
where the sum is not taken over the index $ k $.
This transformation matrix may also be expressed
in terms of the Laplace transform $ {\cal L} $
\cite{OMY96} as
\begin{equation}
{\cal U}(t) = {\cal L}^{-1} [ ( s {\bf 1} + i {\cal H} )^{-1} ] .
\end{equation}
The coefficients $ g^k_{ij} $ are readily found
by using the Laplace transform as
\begin{equation}
g^k_{ij} = \lim_{s \rightarrow - i \lambda_k}
( s + i \lambda_k ) ( s {\bf 1} + i {\cal H} )_{ij}^{-1} ,
\end{equation}
where in the leading orders of $ \kappa $
\begin{eqnarray}
& & ( s {\bf 1} + i {\cal H} )^{-1}
\nonumber \\
& & \ \ \simeq \Lambda^{-1} (s) \left( \begin{array}{ccc}
s_a s_c \ \ & - i \Omega_N s_a \ \ & - \kappa \Omega_N \\
- i \Omega_N s_a \ \ & s_a s_b \ \
& -i \kappa s_b \\
- \kappa \Omega_N \ \ & -i \kappa s_b \ \
& s_c s_b + \Omega_N^2
\end{array} \right)
\label{eqn:siH}
\end{eqnarray}
with
\begin{equation}
s_i \equiv s + i \omega_i + \frac{1}{2} \gamma_i ,
\end{equation}
and
\begin{eqnarray}
\Lambda (s) &=& {\rm det}( s{\bf 1} + i {\cal H} )
\nonumber \\
&=& ( s + i \lambda_1 )( s + i \lambda_2 )( s + i \lambda_3 ) .
\end{eqnarray}
The respective particle numbers are then obtained
with the initial value
$ {\cal N}(0) = {\rm diag.} ( 0 , {\bar n}_c , {\bar n}_a ) $ as
\begin{equation}
n_i (t) = {\cal N}_{ii} (t)
= r_{ic} (t) {\bar n}_c + r_{ia} (t) {\bar n}_a ,
\end{equation}
where
\begin{equation}
r_{ij}(t) = \sum_{m,n} g_{ij}^{m*} g_{ij}^n
\left[ \left( 1 - \frac{\gamma_j}{\Lambda_{mn}} \right)
{\rm e}^{- \Lambda_{mn} t}
+ \frac{\gamma_j}{\Lambda_{mn}} \right]
\end{equation}
with
\begin{equation}
\Lambda_{mn} = - i ( \lambda_m^* - \lambda_n ) .
\end{equation}

The eigenmodes of $ {\cal H} $ are readily
found as follows in the limit of $ \kappa \rightarrow 0 $,
which is indeed sufficient for $ \kappa / \gamma < 10^{-15} $ or so:
\begin{eqnarray}
\lambda_1 &=& \omega_b - \frac{i}{2} \gamma_b - \Delta_{bc}
+ \left[ ( \Delta_{bc} )^2 + \Omega_N^2 \right]^{1/2} , \\
\lambda_2 &=& \omega_c - \frac{i}{2} \gamma_c + \Delta_{bc}
- \left[ ( \Delta_{bc} )^2 + \Omega_N^2 \right]^{1/2} , \\
\lambda_3 &=& \omega_a - \frac{i}{2} \gamma_a ,
\end{eqnarray}
where $ \Delta_{bc} \equiv \frac{1}{2} ( \omega_b - \omega_c )
- \frac{i}{4} ( \gamma_b - \gamma_c ) $,
and $ z^{1/2} \equiv | z |^{1/2} \exp [ i \arg (z) / 2 ] $.
The eigenmodes $ \lambda_1 $ and $ \lambda_2 $,
which mainly consist of the atom and photon,
are determined by the equation
\begin{equation}
\left( \lambda - \omega_b + \frac{i}{2} \gamma_b \right)
\left( \lambda - \omega_c + \frac{i}{2} \gamma_c \right) - \Omega_N^2 = 0 .
\end{equation}
The eignemode $ \lambda_3 $ is, on the other hand,
almost identical with the axion mode.
These eigenmodes of $ {\cal H} $ satisfy the conditions
\begin{equation}
{\rm Re} [ \Lambda_{mn} ]
= - ( {\rm Im} [ \lambda_m ] + {\rm Im} [ \lambda_n ] ) > 0 ,
\end{equation}
so that the respective particle numbers $ n_i (t) $
converge to certain asymptotic values for $ t \rightarrow \infty $.

\section{Langevin picture}
\label{appendix:langevin}

We here describe the quantum evolution of the axion-photon-atom system
in the Langevin picture
\cite{OMY96}, which is supplementary
to the treatment in the Liouville picture reproducing the same results
for the quantum averages of the relevant quantities.
The Langevin equation for the damped oscillators is given by
\begin{equation}
\frac{ d q_i}{dt}
= -i {\cal H}_{ij} (t) q_j + F_i .
\label{eqn:dq/dt}
\end{equation}
The external forces $ F_i $ are introduced
for the Liouvillian relaxations.
These external operators obey the following correlation property
\cite{Louisell90}:
\begin{eqnarray}
\langle F_i^{\dagger} ( \tau ) F_j ( \tau^\prime ) \rangle
&=& \delta_{ij} \gamma_i {\bar n}_i \delta ( \tau - \tau^\prime ) ,
\nonumber \\
\langle F_i^{\dagger} ( \tau ) q_j (0) \rangle &=& 0 \
( \tau , \tau^\prime > 0 ) .
\label{eqn:correl-F}
\end{eqnarray}

The solution of the Langevin equation (\ref{eqn:dq/dt}) is readily found
with the transformation matrix $ {\cal U}(t) $
representing the time evolution due to $ {\cal H} (t) $,
as given in Eq. (\ref{eqn:Ut}):
\begin{equation}
q_i (t) = {\cal U}_{ij} (t) q_j (0)
+ \int_0^t \left[ {\cal U}(t) {\cal U}^{-1} ( \tau ) \right]_{ij}
F_j ( \tau ) d \tau .
\label{eqn:q(t)sol}
\end{equation}
Then, the time evolution of the particle numbers is calculated
with this solution (\ref{eqn:q(t)sol}) of the Langevin equation
and the correlations (\ref{eqn:correl-F}) as
\begin{eqnarray}
& & {\cal N}_{ij} (t) = \langle q_i^\dagger (t) q_j (t) \rangle
\nonumber \\
& & \ \ = {\cal U}^*_{ik} (t) \left[ {\cal N}(0)
+ \int_0^t {\cal U}^{-1 *} ( \tau ) {\cal D} {\cal U}^{-1 {\rm T}} ( \tau )
d \tau \right]_{kl} {\cal U}^{\rm T}_{lj} (t) ,
\label{eqn:N(t)}
\end{eqnarray}
where $ {\cal D}_{ij} = \gamma_i {\bar n}_i \delta_{ij} $.
This solution for $ {\cal N} (t) $ just coincides
with that given in Eq. (\ref{eqn:Ntsol})
which is obtained in the Liouville picture.

\newpage

\vspace*{1cm}

\newpage

\vspace*{1cm}

\newpage

\vspace*{1cm}

\newpage

\vspace*{1cm}

\newpage

\vspace*{1cm}

\newpage

\vspace*{1cm}

\newpage

\vspace*{1cm}

\newpage

\vspace*{1cm}

\newpage

\vspace*{1cm}

\newpage

\vspace*{1cm}

\newpage

\vspace*{1cm}

\newpage

\vspace*{1cm}

\end{document}